\documentclass[prl,a4paper,twocolumn,superscriptaddress,floatfix,nofootinbib,showpacs]{revtex4-1}
\pdfoutput=1
\usepackage{graphicx} % includes figure files
\usepackage{amsmath} \usepackage{array} \usepackage{amssymb}
\usepackage{amsfonts} \usepackage{color} \usepackage{enumitem}
\usepackage{bm} \usepackage{verbatim} \usepackage{hyperref}
\usepackage{morefloats} \usepackage{dcolumn} \usepackage{xspace}
\usepackage[T1]{fontenc} %\usepackage{natbib} %\usepackage{wrapfig}  
\usepackage{relsize} \usepackage{footnote} \usepackage{lipsum}

\hypersetup{
    citecolor=blue,
    colorlinks=true,
    urlcolor=blue
}

\newcommand{\vektor}[1]{{\bf #1}} 
 
\newcommand{\la}{\langle} \newcommand{\ra}{\rangle}
 
\newcommand{\be}{\begin{equation*}} \newcommand{\ee}{\end{equation*}}
\newcommand{\bea}{\begin{eqnarray*}} \newcommand{\eea}{\end{eqnarray*}}

\newcommand{\specialcell}[2][c]{\begin{tabular}[#1]{@{}c@{}}#2\end{tabular}}
 \newcommand{\fref}[1]{Fig.~\ref{#1}} 
 \newcommand{\tref}[1]{Table \ref{#1}}

\makeatletter
\def\blfootnote{\xdef\@thefnmark{}\@footnotetext}
\makeatother

\newcommand{\sector}[1]{\mbox{$\hat{\mathrm{#1}}$}}

\newcommand{\etal}{\emph{et al.}}

\begin{document}

\title{Symmetry fractionalization in the topological phase of the 
  spin-$\frac{1}{2}$ $J_1$-$J_2$ triangular Heisenberg model}

\author{S. N. Saadatmand} \email{s.saadatmand@uq.edu.au}
\affiliation{ARC Centre for Engineered Quantum Systems, School of Mathematics and Physics,
  The University of Queensland, St Lucia, QLD 4072, Australia}
\author{I. P. McCulloch}
\affiliation{ARC Centre for Engineered Quantum Systems, School of Mathematics and Physics, 
  The University of Queensland, St Lucia, QLD 4072, Australia}

\begin{abstract}

        Using density-matrix renormalization-group calculations for infinite cylinders, we elucidate the
        properties of the spin-liquid phase of the spin-$\frac{1}{2}$ $J_1$-$J_2$ Heisenberg model
        on the triangular lattice. We find \emph{four} distinct ground-states characteristic of a 
        non-chiral, $Z_2$ topologically ordered state with vison and spinon excitations.
        We shed light on the interplay of topological ordering and global symmetries in the model 
        by detecting fractionalization of time-reversal and space-group dihedral symmetries in the anyonic 
        sectors, which leads to coexistence of symmetry protected and intrinsic topological order.
        The anyonic sectors, and information on the particle statistics, can be characterized by degeneracy 
        patterns and symmetries of the entanglement spectrum. We demonstrate the ground-states on 
        finite-width cylinders are short-range correlated and gapped; however some features in the 
        entanglement spectrum suggest that the system develops gapless spinon-like edge excitations in 
        the large-width limit.

\end{abstract}

\pacs{03.65.Vf, % Phases: geometric; dynamic or topological
      05.30.Pr, % Fractional statistics systems (anyons, etc.)
      71.10.Pm, % Fermions in reduced dimensions (anyons, composite fermions, Luttinger liquid, etc.)
      75.10.Jm, % Quantized spin models, including quantum spin frustration  
      75.10.Kt, % Quantum spin liquids, valence bond phases and related phenomena
      75.10.Pq, % Spin chain models 
      75.40.Mg} % Numerical simulation studies

\date{\today}

\maketitle

%\section{Introduction}
%\label{sec:intro}

%% Paragraph1: topological phases and their general connections to symmetries %%
\emph{Introduction.\textemdash}Topological phases\cite{Wen02,Wen07_book,Chen10} 
are an intriguing form of the quantum matter, which have been 
challenging theorists for the last two decades. Before then, it 
was believed that Landau symmetry-breaking theory\cite{symmetry-breaking} can explain ordering and phase 
transitions of matter through (spontaneous) breaking of a Hamiltonian 
symmetry. However, topological phases can preserve all symmetries and still acquire 
a finite energy gap. Topological phases fall into two broad categories, ``intrinsic 
topological order''\cite{Chen10} on $D \ge 2$ dimensional lattices, and ``symmetry 
protected topological'' (SPT)\cite{Gu09,PollmannSPT} order, which can also exist
in one dimension. For the former phase, there is no local unitary transformation to smoothly 
deform the state into a product state without passing through a phase transition, regardless of the existence 
of symmetries. The canonical example of an intrinsic topological order is the $Z_2$ ground-state of the 
\emph{toric code}\cite{ToricCode}. On the other hand, SPTs are undeformable into 
product states only if protected by a symmetry. The best studied example is 
surely the Haldane phase of odd-integer spin chains\cite{Gu09,PollmannSPT}, including 
the ground-state of the exactly-solved Affleck-Kennedy-Lieb-Tasaki (AKLT)\cite{AKLT} model. 
A key breakthrough was the realization that anyonic statistics associated with intrinsic topological order
corresponds to fractionalization of symmetry. Therefore when 
intrinsic topological order is coupled with lattice symmetries,
the symmetries themselves fractionalize and lead to SPT ordering\cite{SET-equivalents,He14,Zaletel15}, 
which is readily detectable in many numerical methods.

%% Paragraph2: THM, its symmetries and topological ordering %%
In 1973, Anderson\cite{Anderson73} conjectured that  
the spin-$\frac{1}{2}$ triangular Heisenberg model (THM) with antiferromagnetic 
nearest-neighbor (NN) bonds should stabilize a resonating-valance-bond (RVB) ground-state. 
The failure of analytic and numerical studies\cite{no-RVB,OrderedPhases,Li15,Saadatmand15}
to find such a state motivates the search for a minimal extension that increases the
frustration with a next-nearest-neighbor (NNN) term. 
The Hamiltonian is defined as
\begin{equation}
  \label{eq:Hamiltonian}
  H = J_1 \sum_{\langle i,j \rangle} \vektor{S}_i . {\bf S}_j + 
   J_2 \sum_{\langle\langle i,j \rangle\rangle} {\bf S}_i . {\bf S}_j \; ,
\end{equation}
where $\langle i,j \rangle$ ($\langle\langle i,j \rangle\rangle$) indicates the sum over 
all NN (NNN) bonds. We set $J_1=1$ as the unit of the energy henceforth. Previous 
numerical studies using a range of techniques
\cite{Manuel99,Mishmash13,Kaneko14,Zhu-White_DMRG,Hu-Sheng_DMRG,Li15,Iqbal16,Wietek16}
have suggested a spin liquid\cite{Wen02,Chen10} (SL) region, with phase boundaries in the range of
$J_2^{low} \approx 0.05$\cite{Mishmash13} up to $J_2^{high} \approx 0.19$\cite{Manuel99}.
Employing finite-size density matrix renormalization group (DMRG)\cite{OriginalDMRG,Schollwock11} 
and using fixed aspect-ratio scaling of magnetic order parameters, 
we find phase boundaries of $0.101(4) \leqslant J_2 \leqslant 0.136(4)$, the calculation of
which will be described in more depth in a future work\cite{Saadatmand-prep}. In this Letter,
we focus on the properties of the SL phase itself.
For classical spins, the model has a phase transition
at $J_2 = 0.125$ between two magnetically ordered phases\cite{OrderedPhases}.
This point roughly coincides with the center of the spin-liquid region for the quantum model,
and in this work we focus on $J_2 = 0.125$. 
While there is nothing forbidding coexistence of spontaneous symmetry breaking and topological order,
the Hastings-Oshikawa-Lieb-Schultz-Mattis theorem\cite{HastingsTheorem} in two-dimensions states that
the absence of symmetry breaking in a spin-$\frac{1}{2}$ system on even-width cylinders implies
that the ground-state is a SL with either gapless (algebraic) excitations, or 
gapped with degenerate ground-states and anyonic excitations.
Thus absence of symmetry breaking is a sufficient (but not necessary) condition for a SL.
Previous DMRG studies\cite{Zhu-White_DMRG,Hu-Sheng_DMRG} have argued for a gapped $Z_2$ toric-code 
SL, and have obtained two possible ground-states by the presence (absence) of 
free spins near the boundaries of finite cylinders.
However the properties of these states are unclear, since depending on the sector chosen,
the state may develop chiral order\cite{Hu-Sheng_DMRG},
or breaking of ${\bf C}_6$ rotational symmetry\cite{Zhu-White_DMRG,Hu-Sheng_DMRG,SupplementalMaterials} 
leading to a nematic SL. Recent studies\cite{Qi15,Zaletel15,Zaletel-PRL} focused on the 
kagome lattice show that the time-reversal symmetric $Z_2$ SL can be fully characterized
by the symmetry properties of lattices on tori or infinite cylinders via the 
projective symmetry group (PSG) classifications\cite{Wen02,Huh11,Essin13,Qi15,Zaletel15}.

% \section{Method}
% \label{sec:model}

%%%%%%%%%%%%%%%%%%%%%%%%%%%%%%%%%%%%%%%%%%%%%%%%%%%%%%%%%%%%%%%%%%%%%%%%%%%
\begin{figure}
  \begin{center}
    \includegraphics[width=0.99\columnwidth]{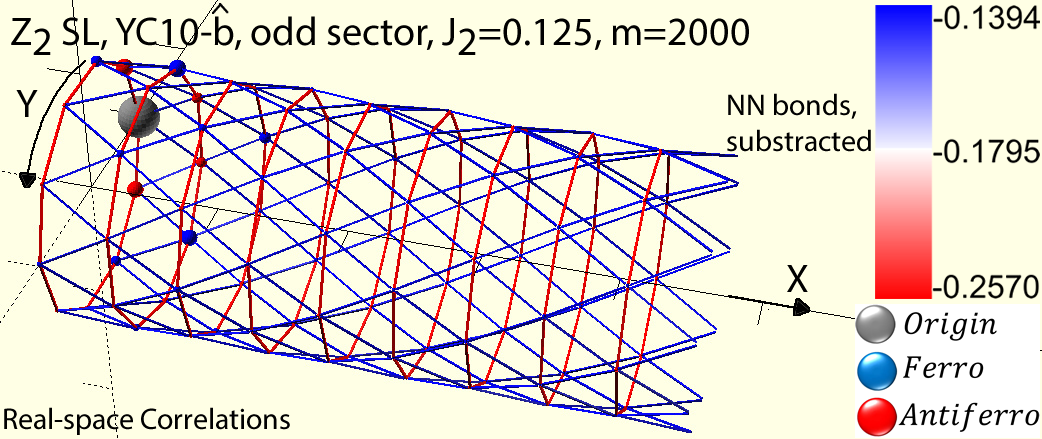}
    \caption{(Color online) 
    Visualization of the triangular lattice on an infinite 
    YC structure. The size and color of the spheres indicate the long-range correlation with the 
    principal (grey) site. The color of the bonds indicates the strength of the NN correlations.
    The average of NN correlation is subtracted from each bond to highlight the anisotropy pattern. 
    \label{fig:LatticeVisualisation}}
  \end{center}
\end{figure}
%%%%%%%%%%%%%%%%%%%%%%%%%%%%%%%%%%%%%%%%%%%%%%%%%%%%%%%%%%%%%%%%%%%%%%%%%%% 

%% Paragraph1: lattice geometry, the control parameter of interest, details on MPS and DMRG method on infinite cylinders %%
\emph{Method.\textemdash}We consider a triangular lattice structure that is wrapped around an infinite
cylinder. We employ the infinite Matrix Product States 
(iMPS)\cite{MPS,McCulloch08,Schollwock11} ansatz, via the infinite DMRG (iDMRG) 
algorithm\cite{McCulloch08,Schollwock11} with single-site optimization\cite{Hubig15} and utilizing $SU(2)$ symmetry 
to obtain translationally-invariant variational ground-states on an infinite cylinder. 
We keep up to $m=5,000$ states, approximately equivalent to 15,000 states of a $U(1)$-symmetric basis.  
We use the so-called YC structure, where the infinite-length cylinder has a circumference equal 
to the number of sites in Y-direction, $L_y$. The mapping of the MPS chain on the cylinder is set 
to minimize the one-dimensional range of NN and NNN interactions. The setup is shown in
\fref{fig:LatticeVisualisation}, indicating also the correlations of a typical ground-state. 
Bipartite quantities, e.g. reduced density matrix ($\rho_r$) 
and entanglement entropy\cite{SupplementalMaterials}, are measured 
by defining a Y-direction \emph{cut} through the cylinder without crossing any vertical bond.
The framework of iDMRG is a natural candidate for calculating symmetry properties, 
since excitations can be introduced at cylinder 
edges by manipulating the symmetries of the wavefunction. Unlike the case for finite systems,
the `edges' are effectively at infinity, so they do not affect the translation symmetry of the wavefunction.
The $Z_2$ SL of the RVB-type carries vison excitations, and bosonic and fermionic spinons\cite{Senthil00}.
We control the even/odd parity of spinon flux in the ground-state by setting $SU(2)$ 
quantum-numbers (global spins, $S$) to be either integers (even-sector -- no spinon) or 
odd-half-integers (odd-sector -- with spinons) at the unit cell boundary. We cannot directly control 
the vison flux through the cylinder, so we can only obtain two ground-states
for each cylinder geometry. However for finite-$L_y$ cylinders the degeneracy is expected to be lifted,
and fortunately we find that the ground-states for different width cylinders also give the
vison and non-vison sectors, allowing us to obtain all four combinations of even/odd spinons 
and presence/absence of a vison flux. We note that Metlitski and Grover\cite{Metlitski11} 
and Kolley \emph{et al.}\cite{Kolley13} established the observation of 
a Tower of States (TOS) in the low-lying part of the entanglement 
spectrum (ES)\cite{OriginalES,Kolley13} as a ``smoking gun'' evidence for 
the existence of magnetically ordered states (carrying Nambu-Goldstone excitations). 
We confirm the non-magnetic nature of the phase by the absence of TOS in the ES, 
regardless of the anyon sector (see below and also Ref.\ \onlinecite{Saadatmand-prep}).

%% Paragraph2: elementary excitations on tori/infinite cylinders and connections to symmetry groups + PSGs
%% of a single anyon + how we classify sectors using MPS/iDMRG? %%
We obtain a structure of four anyon sectors of the $Z_2$ toric-code-type topological order\cite{Kitaev06}
that comprises the identity \sector{i}-anyon (carries no spinon or vison flux), 
a bosonic spinon \sector{b}-anyon (carries a $S=\frac{1}{2}$ spin), a \sector{v}-anyon (carries
a vison and has a $\pi$-flux threading the cylinder, equivalent to possessing anti-periodic 
boundary conditions in the Y-direction), and finally a fermionic spinon 
\sector{f}-anyon (a composite excitation, which carries both a $S=\frac{1}{2}$ spin and 
a $\pi$-flux). In this letter, we work in a minimally entangled states
(MES)\cite{Zhang12,Jiang12} basis introduced in Refs.\ \onlinecite{Zaletel-PRL,Zaletel15}
for the four-dimensional ground-state manifold and preserves SPT ordering.
For even-$L_y$ cylinders, each unique MES state\cite{Zaletel15} corresponds to threading an anyonic flux 
in the long-direction and creating a particle/anti-particle pair of $\sector{a}$ at infinity, namely 
$| U_{L_y}^{a/\tilde{a}} \ra$ (also denoted as YC$L_y$-\sector{a} sector). Given a particular MES, 
the action of a global symmetry group ($\bar{g}$) member, $\Gamma_g$, on the state can be considered as two 
independent actions on each anyon, i.e. $\Gamma_g | U_{L_y}^{a/\tilde{a}} \ra = \Upsilon_g | U_{L_y}^a \ra 
\otimes \Upsilon_{\tilde{g}} | U_{L_y}^{\tilde{a}} \ra$, where $\Upsilon_g$'s are 
unitary operators acting on a single anyon $| U_{L_y}^a \ra$. 
Anyons can fractionalize\cite{Essin13,Qi15,Zaletel-PRL,Zaletel15} the symmetry, 
$\bar{g}$, by factorizing an identity member of the group (square root of $\bar{g}$). $\bar{g}$ is always a 
\emph{linear} representation (it is describing a physical symmetry), but $\Upsilon_g$'s can now 
form a non-trivial PSG, which is a central extension of the original group\cite{Wen02}. 
In the MPS representation of the ground-state, the $\Upsilon_g$ can be expressed as operators
acting on the `auxiliary' basis, i.~e. the basis of the entanglement Hamiltonian, $-\ln(\rho_r)$, on a bipartite cut. 
Thus the existence of a PSG through measurements of $\Upsilon_g$'s 
implies 1D SPT ordering\cite{Zaletel15}, by considering rings as single ``super-sites'' 
(global symmetries along Y-direction are now internal symmetries when viewed as a 1D chain), which
is straightforward to detect using iMPS techniques\cite{PollmannPSG}.

% \section{Ground-state Energies}
% \label{sec:energy}

%%%%%%%%%%%%%%%%%%%%%%%%%%%%%%%%%%%%%%%%%%%%%%%%%%%%%%%%%%%%%%%%%%%%%%%%%%%
\begin{figure}
  \begin{center}
    \includegraphics[width=0.99\columnwidth]{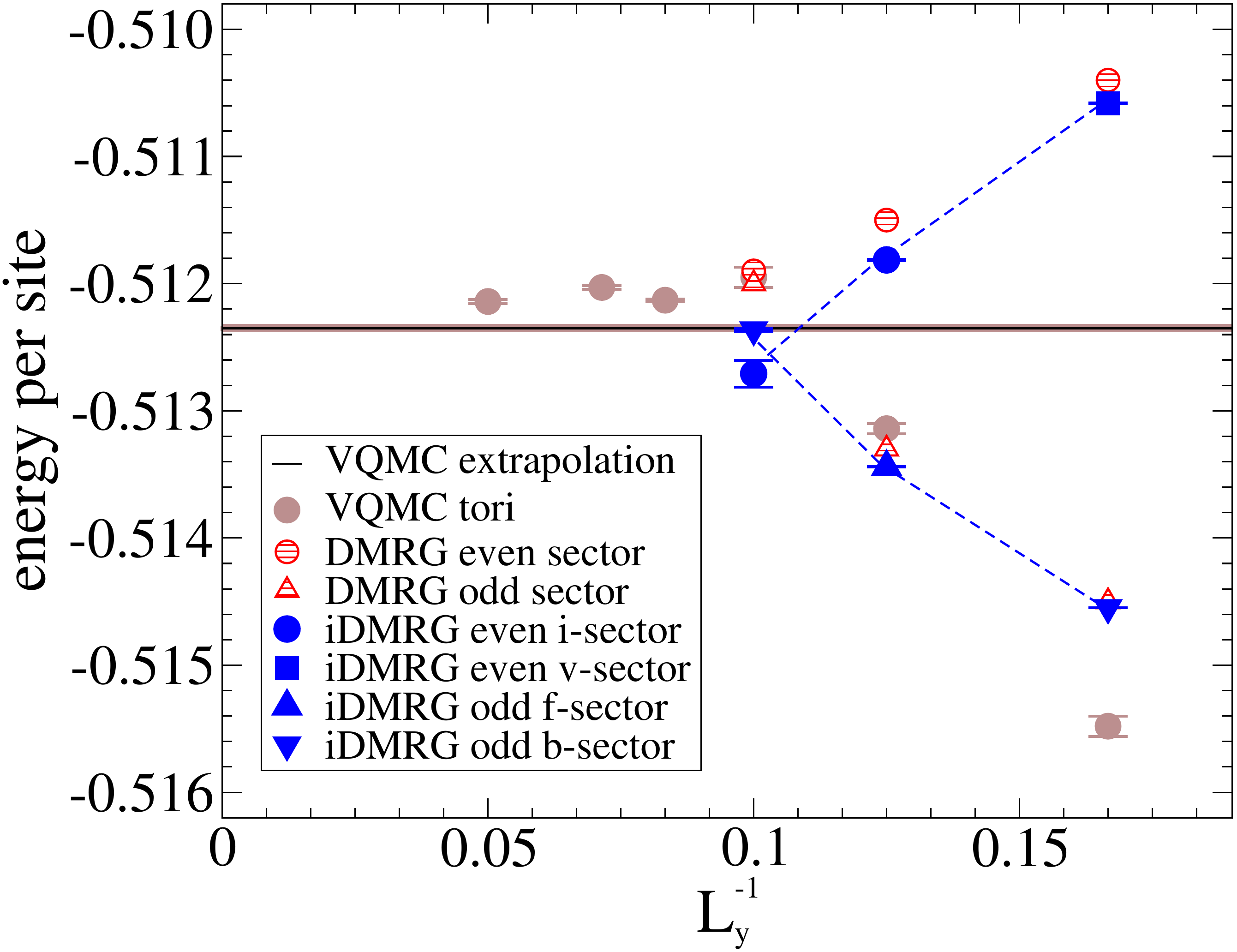}  
    \caption{(Color online) 
    Ground-state energies at $J_2=0.125$ against inverse cylinder width. Our results for
    infinitely long cylinders with extrapolation to zero variance are in blue.
    Dashed lines are guides to the eye. Red shaded symbols are finite-size DMRG 
    results from Ref.\ \onlinecite{Hu-Sheng_DMRG}. Brown symbols are variational quantum Monte-Carlo 
    (VQMC) results on $L \times L$ tori with the horizontal
    line indicating an $L \rightarrow \infty$ extrapolation, from Ref.\ \onlinecite{Iqbal16}.
    \label{fig:energy}}
  \end{center}
\end{figure}
%%%%%%%%%%%%%%%%%%%%%%%%%%%%%%%%%%%%%%%%%%%%%%%%%%%%%%%%%%%%%%%%%%%%%%%%%%%

%% Paragraph1: extrapolated ground-state energies %%
\emph{Ground-state energies.\textemdash}We present ground-state energies 
of anyonic sectors in \fref{fig:energy}. Energies are extrapolated to the thermodynamic limit 
of basis size $m \rightarrow \infty$, using a linear fit against the
energy variance per site (see Ref.~\onlinecite{SupplementalMaterials} for details). 
We suggest that fitting against variances is the most accurate method for 
extrapolating energies in DMRG (more reliable than
extrapolation with respect to the DMRG truncation error). The different
topological sectors are expected to acquire slightly different energies on finite-width 
cylinders. Depending on $L_y$, we find that the actual ground-state in the even/odd sectors
varies as to whether it contains a $\pi$ vison flux or not. In some cases, especially
for smaller widths, we have been able to construct
variational wavefunctions in the other sectors by manipulating the wavefunction 
(i.~e. to force a particular symmetry state), but the resulting states are rather 
unstable and have considerably higher energies. However the overall 
behavior of energies indicates that the difference between energy of even- and odd-sectors
is rapidly decreasing with increasing $L_y$. This is consistent with having a degenerate ground-state in the
thermodynamic limit $L_y\rightarrow\infty$. Interestingly, there is an energy crossover between even/odd-sectors
already for YC10, which makes it unreliable to estimate an energy for $L_y\rightarrow\infty$ 
limit. We note that our energies per site for larger system widths are somewhat 
lower than previously published results.

% \section{Symmetry Group Measurements}
% \label{sec:SymmetryGroups}

%%%%%%%%%%%%%%%%%%%%%%%%%%%%%%%%%%%%%%%%%%%%%%%%%%%%%%%%%%%%%%%%%%%%%%%%%%%%%%%%%%%%%%%%%%%%%%%%%%%%%%%%%%%%%%%%%%%%%%%%%%%%%%%%%%%%%%%%  
\newcommand\T{\rule{0pt}{2.6ex}}
\begin{table}
  \caption{Summary of topological invariants for $L_y \times \infty$ cylinders at $J_2=0.125$.
    $|\la R_y \ra|$, $|\la T_y \ra|$, and $|\la \tau \ra|$ are close 
    to $1.0$ in all cases\cite{SupplementalMaterials}.
  \label{table:SymmetryGroups}}
  \begin{center}
    {%\renewcommand{\arraystretch}{1.25}
     \begin{tabular}{ |c|c|c|r@{$ 0. $}l|r@{$ 1 \; \pm \; $}l|c| }
	\hline \hline 
        \tiny Structure \normalsize & \tiny \specialcell[c]{spin-$\frac{1}{2}$\\boundary} \normalsize \T & \tiny \specialcell[c]{
        degeneracy\\of ES} \normalsize & \multicolumn{2}{c|}{$\la C[D_{L_y}] \ra$} & \multicolumn{2}{c|}{$\la C[\tau^2] 
        \ra$} & \tiny \specialcell[c]{sector} \normalsize \\ \hline
        %%%%%%%%%%%%%%%%%%%%%%%%%%%%%%%%%%%%%%%%%%%%%%%%%%%%%%%%%%%%%%%%%%%%%%%%%%%%%%%%%%%%%%%%%%%%%%%%%%%%%%%%%%%%%%%%%%%%%
        YC6 & even & \scriptsize 2-fold \normalsize & $-$ & $999996$ & & $10^{-11}$ \T & \sector{v} \\ \hline
        %%%%%%%%%%%%%%%%%%%%%%%%%%%%%%%%%%%%%%%%%%%%%%%%%%%%%%%%%%%%%%%%%%%%%%%%%%%%%%%%%%%%%%%%%%%%%%%%%%%%%%%%%%%%%%%%%%%%%
        YC6 & odd & \scriptsize 2-fold \normalsize & & $9999998$ & $-$ & $10^{-14}$ \T & \sector{b} \\ \hline
        %%%%%%%%%%%%%%%%%%%%%%%%%%%%%%%%%%%%%%%%%%%%%%%%%%%%%%%%%%%%%%%%%%%%%%%%%%%%%%%%%%%%%%%%%%%%%%%%%%%%%%%%%%%%%%%%%%%%%
        YC8 & even & \tiny \specialcell[c]{non-\\degenerate} \normalsize & & $99998$ & & $10^{-10}$ \T & \sector{i} \\ \hline
        %%%%%%%%%%%%%%%%%%%%%%%%%%%%%%%%%%%%%%%%%%%%%%%%%%%%%%%%%%%%%%%%%%%%%%%%%%%%%%%%%%%%%%%%%%%%%%%%%%%%%%%%%%%%%%%%%%%%%
        YC8 & odd & \scriptsize 4-fold \normalsize & $-$ & $999990$ & $-$ & $10^{-11}$ \T & \sector{f} \\ \hline
        %%%%%%%%%%%%%%%%%%%%%%%%%%%%%%%%%%%%%%%%%%%%%%%%%%%%%%%%%%%%%%%%%%%%%%%%%%%%%%%%%%%%%%%%%%%%%%%%%%%%%%%%%%%%%%%%%%%%%
        YC10 & even & \tiny \specialcell[c]{non-\\degenerate} \normalsize & & $9996$ & & $10^{-9}$ \T & \sector{i} \\ \hline
        %%%%%%%%%%%%%%%%%%%%%%%%%%%%%%%%%%%%%%%%%%%%%%%%%%%%%%%%%%%%%%%%%%%%%%%%%%%%%%%%%%%%%%%%%%%%%%%%%%%%%%%%%%%%%%%%%%%%%
        YC10 & odd & \scriptsize 2-fold \normalsize & & $9998$ & $-$ & $10^{-9}$ \T & \sector{b} \\
        %%%%%%%%%%%%%%%%%%%%%%%%%%%%%%%%%%%%%%%%%%%%%%%%%%%%%%%%%%%%%%%%%%%%%%%%%%%%%%%%%%%%%%%%%%%%%%%%%%%%%%%%%%%%%%%%%%%%%
	\hline \hline
     \end{tabular} 
    }
  \end{center}
\end{table}
%%%%%%%%%%%%%%%%%%%%%%%%%%%%%%%%%%%%%%%%%%%%%%%%%%%%%%%%%%%%%%%%%%%%%%%%%%%%%%%%%%%%%%%%%%%%%%%%%%%%%%%%%%%%%%%%%%%%%%%%%%%%%%%%%%%%%%%%

%% Paragraph1: expansions of results on major symmetry groups %%
\emph{Symmetry group measurements.\textemdash} 
We present our main symmetry group measurements on different anyonic 
sectors and system sizes in \tref{table:SymmetryGroups}. Considering 
the time-reversal symmetry, $\tau$, we find that $|\la \tau \ra|$ is very close to 1.0 in all sectors, 
indicating that time-reversal symmetry is not broken and therefore the ground-state is non-chiral.
A state carrying a spin-$\frac{1}{2}$ spinon flux can be realized by the action of $\tau$ on 
the auxiliary basis, in the form of $C[\tau^2] = \Upsilon_\tau \Upsilon^*_\tau$.
For \sector{b} and \sector{f} sectors, one expects $\la \Upsilon_\tau \Upsilon_\tau^* \ra = -1$, 
anti-symmetric under time-reversal, which is precisely what 
we observe (these SPTs are also protected by parity reflection\cite{SupplementalMaterials}).
A state that carries a $\pi$ vison flux can be detected by the action of the
cylinder dihedral symmetry group $D_{L_y}$ in the Y-direction. 
The elements of the group are generated by reflection around 
a site or bond\cite{SupplementalMaterials}, $R_y$, and a translation by one lattice site, $T_y$.
The linear and projective representations can be distinguished by the commutation
between $R_y$ and a $\pi$-rotation, $T_y^\pi = (T_y)^{\frac{L_y}{2}}$.
Visons fractionalize $D_{L_y}$, acquiring an effective 
anti-periodic boundary condition in Y-direction, whereby reflection and $\pi$ rotations
anticommute, $R_y T_y^{\pi} = - T_y^{\pi} R_y$. Thus one expects $ C[D_{L_y}] = \la \Upsilon_{R_y} 
\Upsilon_{T^\pi_y} \Upsilon_{R_y}^\dagger \Upsilon_{T^\pi_y}^\dagger \ra = -1$ 
for the \sector{v} and \sector{f} sectors;
i.e.\ $D_{L_y}$ fractionalizes into a PSG with an invariant gauge group\cite{Wen02,Zaletel15} of $Z_2$. 
Combined, the measurements of $C[\tau^2]$ and $C[D_{L_y}]$ give distinct topological
invariants for the four sectors, and imply fusion rules\cite{Kitaev06,SupplementalMaterials} of a $Z_2$ SL.
Furthermore, this gives information about the self-statistics, in particular the
obtained topological invariants are incompatible\cite{Zaletel-PRL} with 
the double-semion topological order\cite{DoubleSemion}, since the semion and anti-semion
are time-reversal partners, but here the two spinon sectors have different 
PSGs so they cannot be interchanged under $\tau$. We also present a more comprehensive 
list of symmetry observables in the Supplementary Material\cite{SupplementalMaterials}.

% \section{Entanglement Spectrum}
% \label{sec:ES}

%%%%%%%%%%%%%%%%%%%%%%%%%%%%%%%%%%%%%%%%%%%%%%%%%%%%%%%%%%%%%%%%%%%%%%%%%%%
\begin{figure}
  \begin{center}
    \includegraphics[width=0.45\columnwidth]{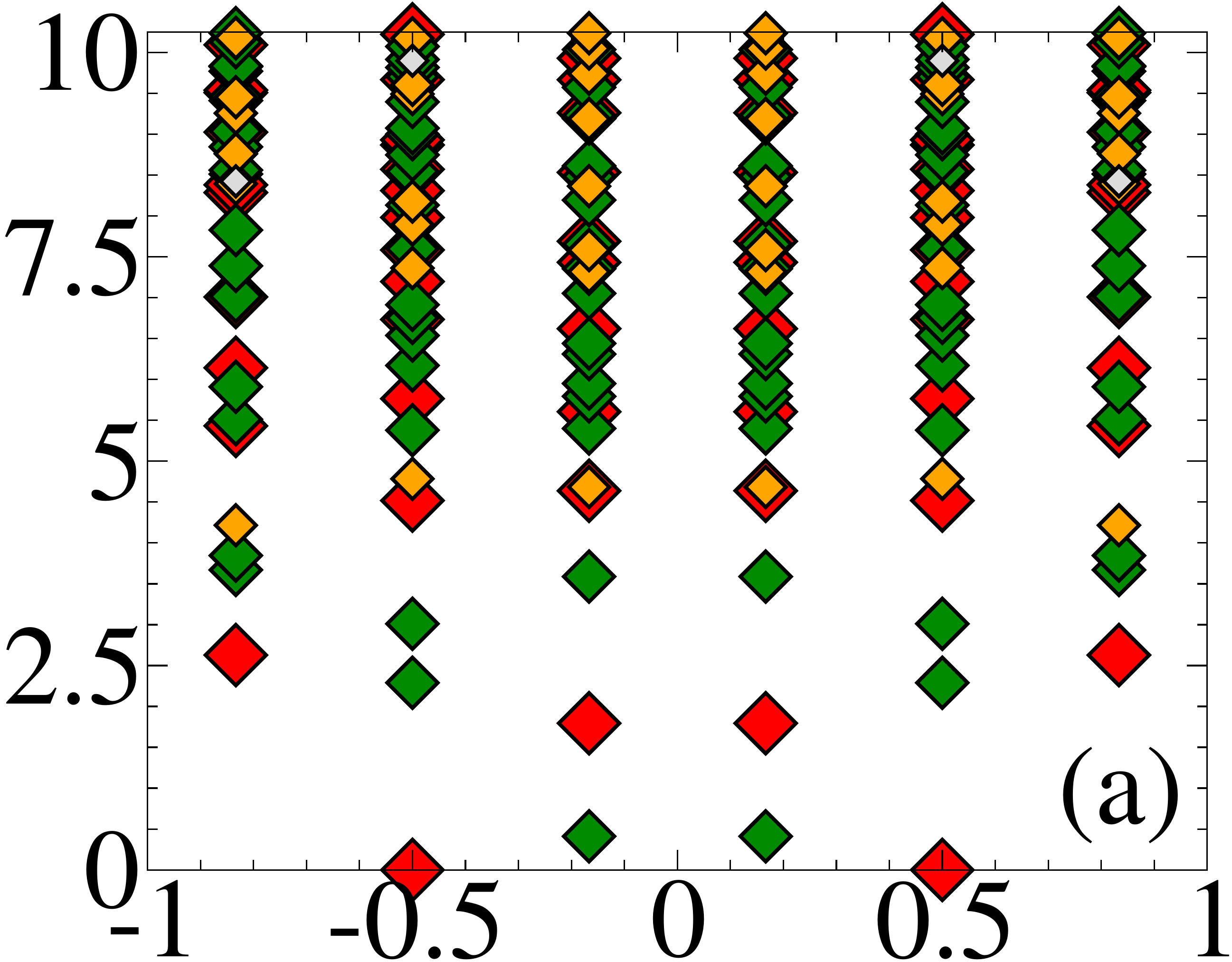}
    \includegraphics[width=0.45\columnwidth]{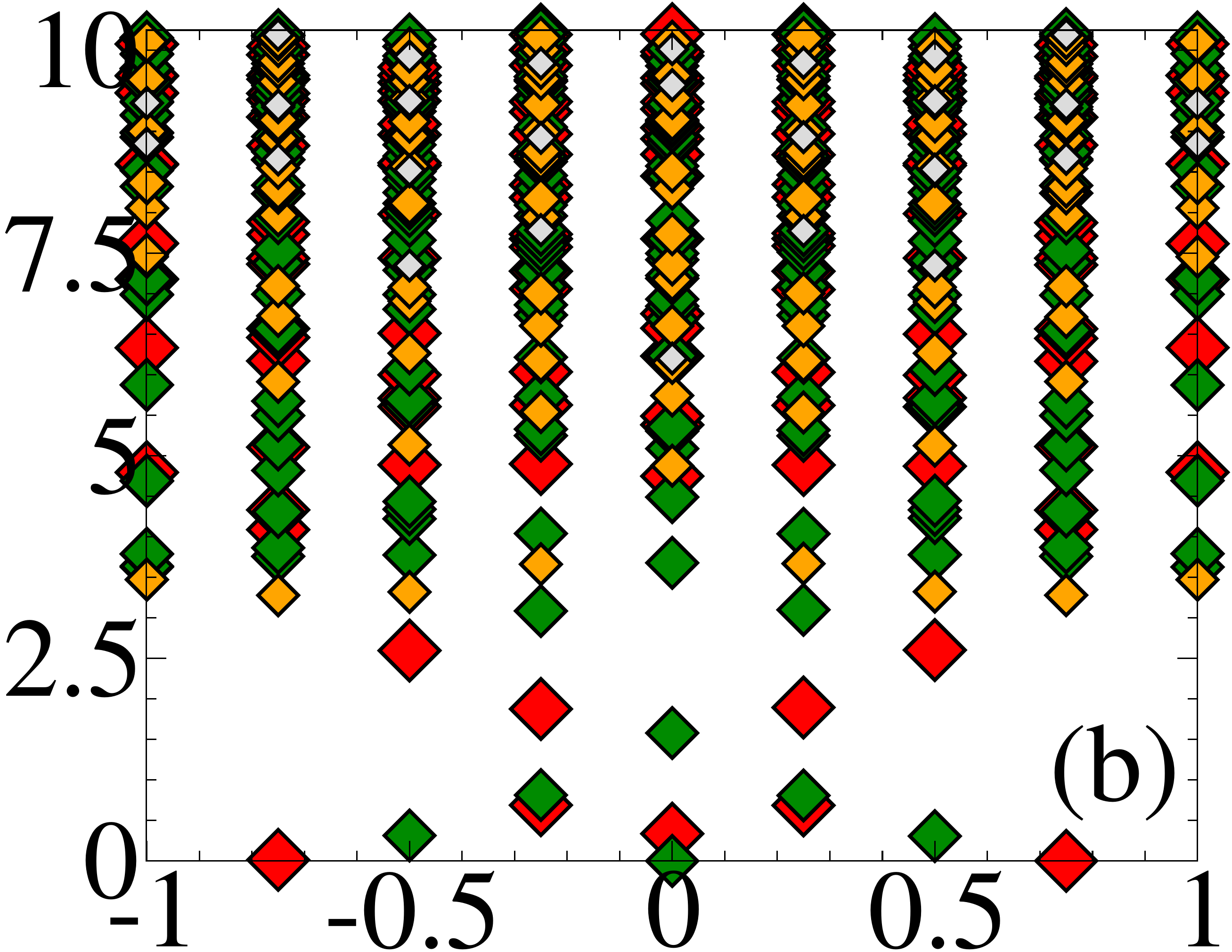} \\
    \includegraphics[width=0.91\columnwidth]{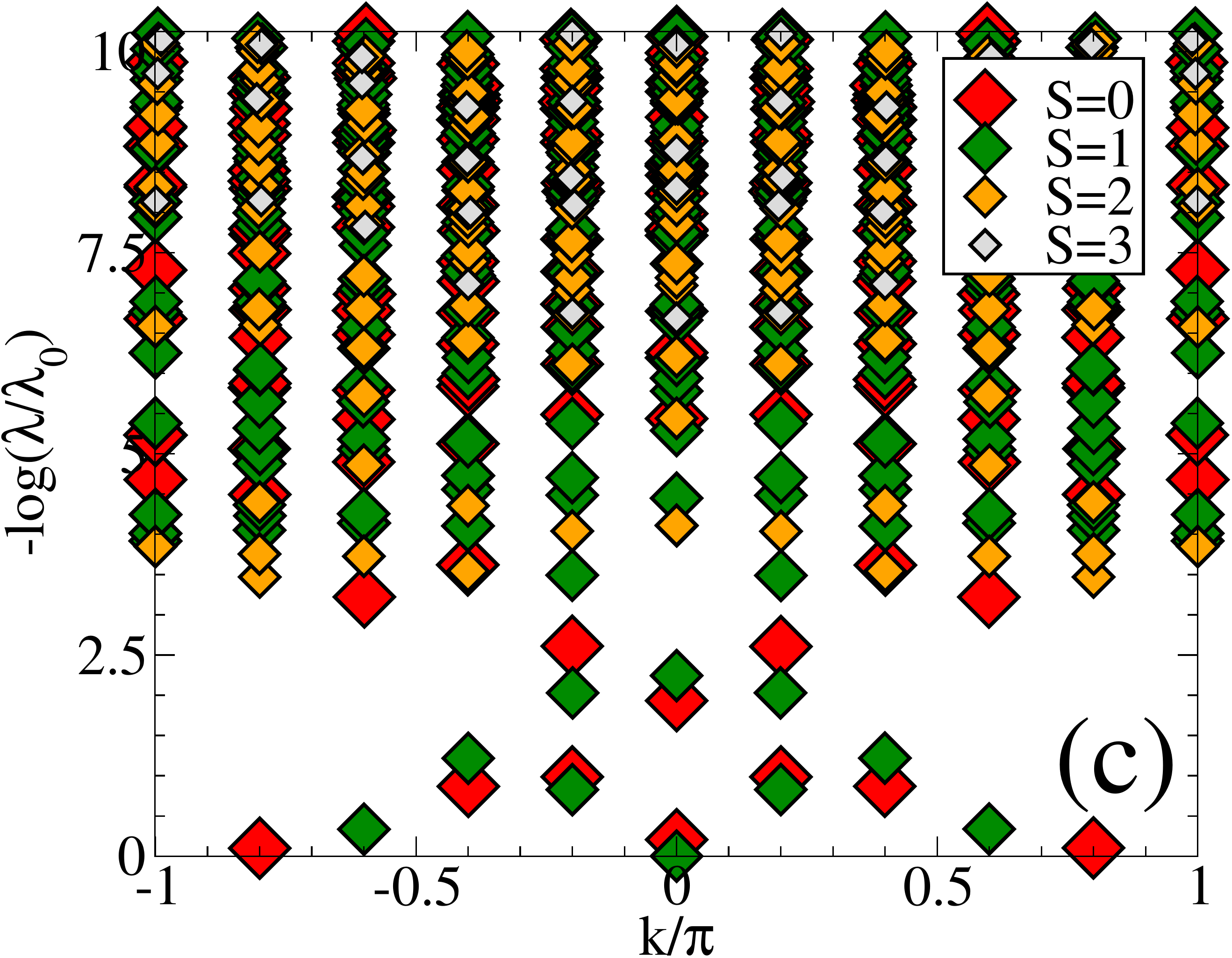}
    \caption{(Color online) 
    The momentum-resolved ES of the \emph{even-boundary} 
    topological sectors for different cylinder circumferences $L_y$, 
    in the spin-liquid region at $J_2=0.125$. The topological sectors are
    (a) YC6-\sector{v}, (b) YC8-\sector{i} and (c) YC10-\sector{i}.
    \label{fig:ES-even}}    
  \end{center}
\end{figure}
%%%%%%%%%%%%%%%%%%%%%%%%%%%%%%%%%%%%%%%%%%%%%%%%%%%%%%%%%%%%%%%%%%%%%%%%%%%

%%%%%%%%%%%%%%%%%%%%%%%%%%%%%%%%%%%%%%%%%%%%%%%%%%%%%%%%%%%%%%%%%%%%%%%%%%%
\begin{figure}
  \begin{center}
    \includegraphics[width=0.45\columnwidth]{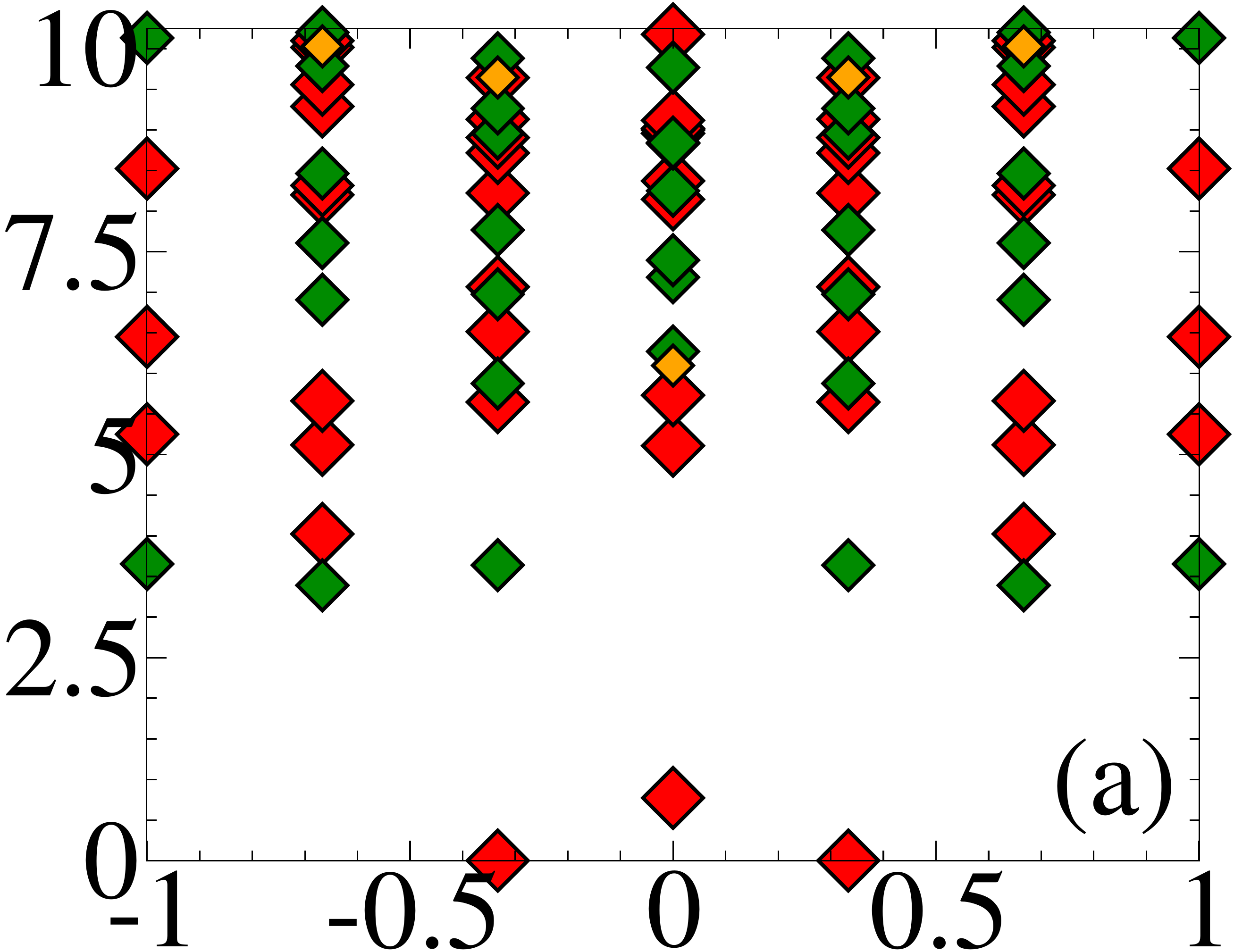}
    \includegraphics[width=0.45\columnwidth]{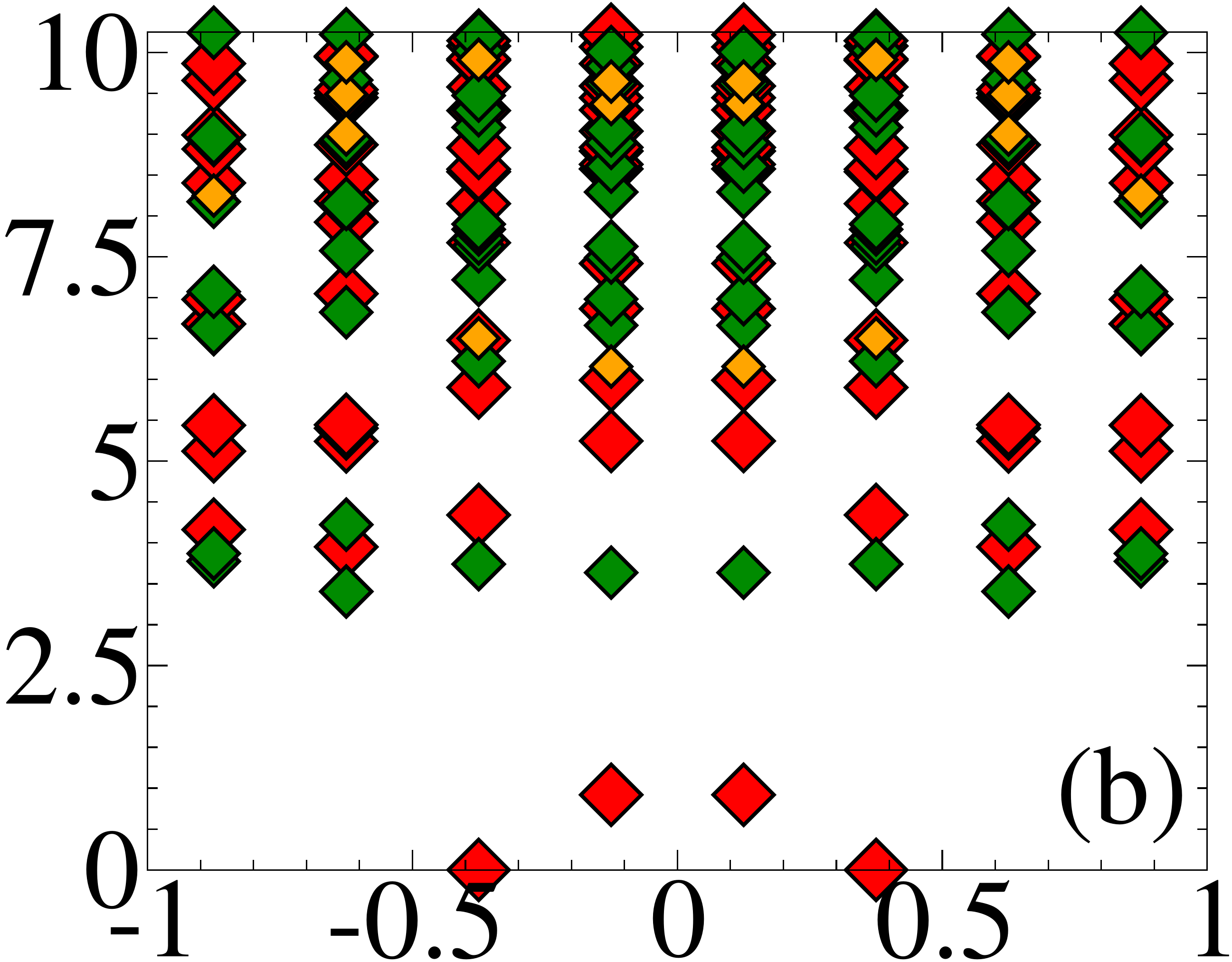} \\
    \includegraphics[width=0.91\columnwidth]{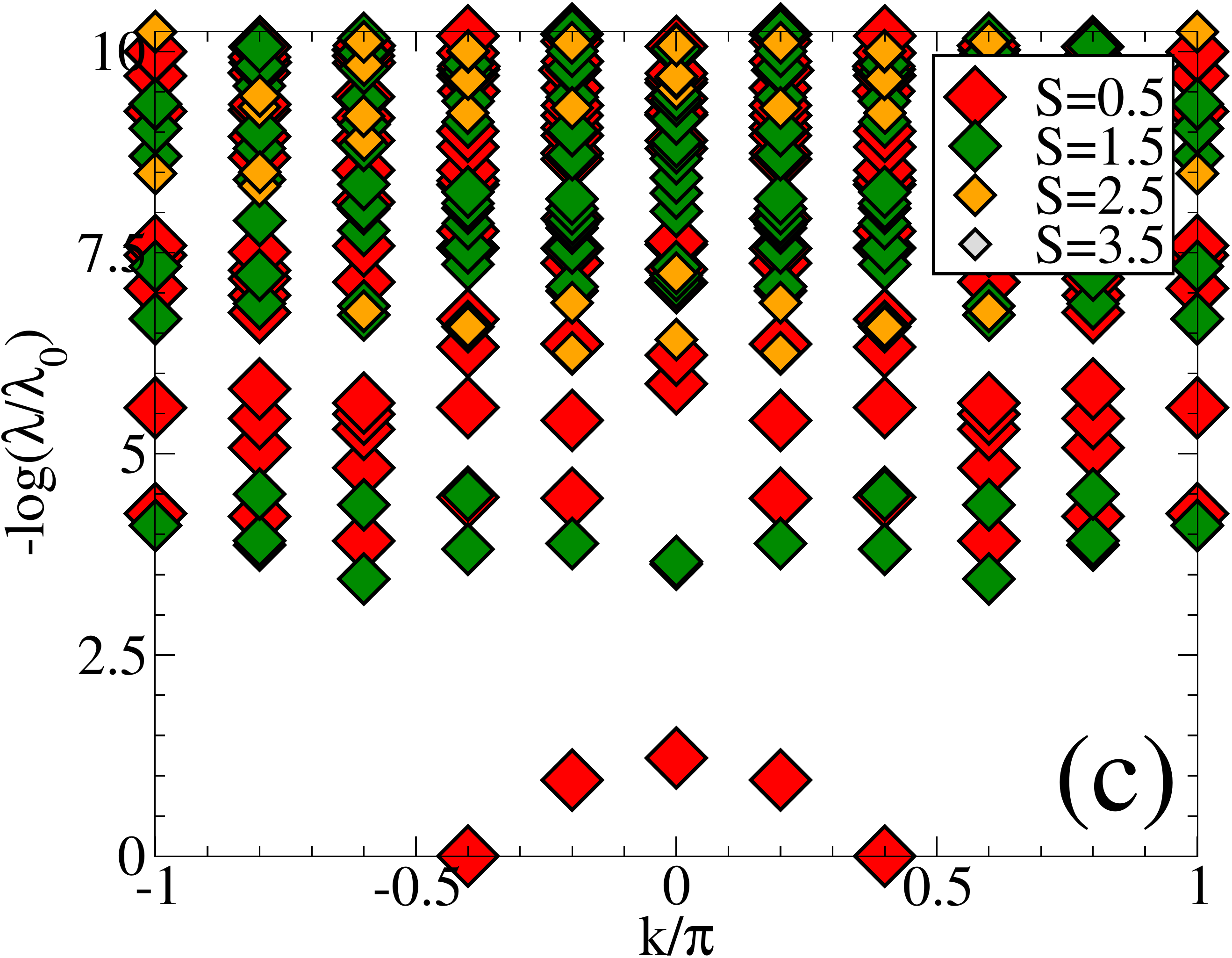}
    \caption{(Color online)
    The momentum-resolved ES of the \emph{odd-boundary} 
    topological sectors for different cylinder circumferences $L_y$, 
    in the spin-liquid region at $J_2=0.125$. The topological sectors are
    (a) YC6-\sector{b}, (b) YC8-\sector{f} and (c) YC10-\sector{b}.
    \label{fig:ES-odd}}    
  \end{center}
\end{figure}
%%%%%%%%%%%%%%%%%%%%%%%%%%%%%%%%%%%%%%%%%%%%%%%%%%%%%%%%%%%%%%%%%%%%%%%%%%%

%% Paragraph1: What is the momentum-resolved ES? %%
\emph{Entanglement spectrum.\textemdash}The ES, set of $\{\lambda_i\}$'s, is a way of 
presenting the eigenvalues of the entanglement Hamiltonian   
analogous to a set of energy levels. $\lambda_i$ can be labeled
by any global symmetry of the system as long as it is preserved in the bipartite cut. 
In this case, we choose $SU(2)$ spin $S$ (preserved explicitly in the calculations),
and $D_{L_y}$, which is \emph{not} preserved exactly; but it is straightforward 
to diagonalize $T_y$ to obtain the momentum-resolved ES.  
In the absence of a $\pi$ vison flux, the allowed $Y$-momenta ($k_n$'s) are 
arranged with a spacing of $k_n = \frac{2\pi n}{L_y}$.
The key difference in the vison sectors is a shift of a half spacing, 
$k_n = \frac{2\pi}{L_y}(n+1/2)$ due to the $\pi$-flux causing an effective 
anti-periodic boundary condition. Because each anyonic sector corresponds to a unique set of 
symmetry group measurements that cannot smoothly deform, sectors have
a uniquely identifying ES (such a unique form of ES 
on infinite cylinders was originally observed in the honeycomb Haldane model\cite{Cincio13}). 
In general it is a non-trivial task to interpret highly-populated ES levels, 
but the overall degeneracy patterns are signatures of SPT ordering, when 
viewing the cylinder as an infinite chain\cite{PollmannSPT,Turner11,Zaletel15}. That is,
in the presence of SPT, every ES state has a multiple of $n$-fold degeneracy, where $n$ is
determined by the symmetry properties of the state. In particular,
the \sector{i}-sector ES has no degeneracy, \sector{b}-sector has 2-fold degeneracy 
associated with half-integer spins (Kramers degeneracy from $C[\tau^2] = -1$), 
\sector{v}-sector has 2-fold degeneracy associated with PSGs of $D_{L_y}$, and the 
\sector{f}-sector has 4-fold degeneracy combining Kramers and PSG of $D_{L_y}$.

%% Paragraph2: what did we observe for momentum-resolved ES of even-sectors? %%
In \fref{fig:ES-even}, we present the ES of even-boundary topological sectors for various
width cylinders. The \sector{v}-sector on YC6 has an exact (up to numerical accuracy) 2-fold degeneracy arising
from $\pm k$ momenta, which is not shared by the \sector{i}-sector (the $k=0$ and $k=\pi$ states
are non-degenerate), which is a proof for the $\pi$-flux.
The low-lying structure is a deformed two-spinon continuum, most easily seen for the larger-width lattices.
We suggest this general pattern (manifested in \fref{fig:ES-even}(c)) is characteristic of 
even sectors and presumably persists in large-$L_y$ limit.   

%% Paragraph3: what did we observe for momentum-resolved ES of odd-sectors? %%
ES results for odd-boundary topological sectors are presented in \fref{fig:ES-odd}. 
The \sector{f}-sector for YC8 has (nearly) 4-fold degeneracy 
and momenta are shifted by $\frac{\pi}{8}$, which indicates
both a spinon and vison $\pi$-flux. The ES of the \sector{b}-sector for YC6 or YC10 is (nearly) 2-fold 
degenerate due to the odd-half-integer spin boundaries, indicating spinons but no $\pi$-flux.
Again the overall pattern of the low-energy structure is consistent between vison and non-vison sectors,
and appears to be converging to a well-defined large-$L_y$ limit.
Intriguingly, the low-energy structure for the odd sector is reminiscent of a Fermi arc\cite{FermiArc},
appearing as an excitation mode that only covers a subset of the Brillouin zone.

%% Paragraph4: ES results in comparison to Ref. [21]'s spectra %%
Hu~\etal\cite{Hu-Sheng_DMRG} presents ES (not momentum-resolved) for two nearly-degenerate, YC8 
ground-states (Fig.~5, corresponding to the \sector{i}- and \sector{f}-sector). Ref.~\onlinecite{Hu-Sheng_DMRG}'s 
\sector{f}-sector ES is consistent with our \fref{fig:ES-odd}(b), however, there is no match between the
\sector{i}-sector spectra. We suggest Ref.~\onlinecite{Hu-Sheng_DMRG}'s 
\sector{i}-sector spectra corresponds to a \emph{chiral state}.

% \section{Discussion}
% \label{sec:discussion}

%% Paragraph1: very short line on what we understood about the phase until now %%
\emph{Discussion.\textemdash}Using $SU(2)$-symmetric iDMRG, we have provided a robust 
demonstration of the properties of the spin-liquid phase of the THM on infinite cylinders, obtaining
\emph{four} ground-states and their ES degeneracy patterns, which we have classified according to their 
symmetry fractionalization properties, consistent to theoretical predictions 
(e.~g. see Ref.~\onlinecite{Zaletel15}). To the best of our knowledge, this is the
first study to observe dihedral symmetry fractionalization in such a model, which shows the 
low-lying structure of THM carries non-chiral $Z_2$ toric-code-type topological order.

%% Paragraph2: clarifying the nature of gap (with the help of observables measured in "supplemental materials") %%
While our calculations are always in the limit of infinite aspect ratio (we do not address directly the 
nature of the 2D limit), we suggest that the degeneracy of the ground-states is robust. 
We are not yet able to directly measure the energy gap to excited states, 
however the iMPS ansatz does readily provide the correlation length, which in all cases
is rather small\cite{SupplementalMaterials}, implying a finite gap for finite-$L_y$. 
However, low-lying structures in the ES contain some interesting features, such as a Dirac cone
(\fref{fig:ES-even}), which will become gapless in the thermodynamic
limit. According to edge-ES correspondence for $Z_2$ topological states enriched by
global symmetries\cite{Ho15}, the system is likely to have gapless edge states. It
is unclear if this would also lead to gapless bulk states, hence we are unable to
rule out the possibility that the system is algebraic SL in the 2D limit.

%% Paragraph3: clarifying the nature of anisotropy (with the help of observables measured in "supplemental materials") %%
In agreement with Hu \emph{et al.}\cite{Hu-Sheng_DMRG}, we 
observe anisotropic ($C_6$-symmetry breaking) correlations
for the odd-sectors only, while the even-sectors appear to get isotropic as the width is increased. 
We were unable to detect the expected topological entanglement entropy
of $-ln(2)$ due to the limited accuracy of the obtained entropy,
and relatively small $L_y$, which is an inherent difficulty with DMRG 
procedure\cite{SupplementalMaterials}. If the system is gapless in the 2D limit, 
then there will be \emph{logarithmic} corrections that make the fit almost impossible to perform 
for numerically accessible system sizes.

%% Paragraph4: possible real-world usages %%
Irrespective of the nature of the state in the 2D limit, we have shown that finite-width
YC structures have short-range correlations and are gapped. A long, narrow cylinder is
a plausible geometry for a quantum-engineered device, and there are recent proposals
for the construction of fermionic Hofstadter-Hubbard model on a cylindrical optical 
lattice\cite{OpticalLattice}. Candidate materials that could be realizations of the $Z_2$ RVB
SL are $\kappa$-(BEDT-TTF)$_2$Cu$_2$(CN)$_3$\cite{OrganicGroup-gapped} 
(with no indication of any gapless spin excitations) 
and EtMe$_3$Sb[Pd(dmit)$_2$]$_2$\cite{OrganicGroup-gapless}
(recognized as a gapless state).

\begin{acknowledgments}
  
  The authors would like to thank Jason Pillay, Henry Nourse,
  Michael Zaletel, Yin-Chen He, Wen-Jun Hu and Shou-Shu Gong for useful discussions.
  The authors would also like to thank Ben Powell for some start-up ideas
  and useful discussions in the early stages of the project.
  This work has been supported by the Australian 
  Research Council (ARC) Centre of Excellence for Engineered Quantum Systems, 
  grant CE110001013. I.P.M also acknowledges support from the 
  ARC Future Fellowships scheme, FT140100625.
  
\end{acknowledgments}

%\bibliography{mybib}

%%%%%%%%%%%%%%%%%%%%%%%%%% Merge with supplemental materials %%%%%%%%%%%%%%%%%%%%%%%%%%%%%
\pagebreak
\widetext
\begin{center}
\textbf{\large Symmetry fractionalization in the topological phase of the spin-$\frac{1}{2}$ $J_1$-$J_2$ triangular Heisenberg model -- 
  Supplemental Material} \\ \bigskip
\normalsize S. N. Saadatmand and I. P. McCulloch \normalsize
\end{center}
%%%%%%%%%% Prefix a "S" to all equations, figures, tables and reset the counter %%%%%%%%%%
\setcounter{equation}{0}
\setcounter{figure}{0}
\setcounter{table}{0}
\setcounter{page}{1}
\makeatletter
\renewcommand{\theequation}{S\arabic{equation}}
\renewcommand{\thefigure}{S\arabic{figure}}
\renewcommand{\bibnumfmt}[1]{[S#1]}
\renewcommand{\citenumfont}[1]{S#1}
%%%%%%%%%%%%%%%%%%%%%%%%%%%%%%%%%%%%%%%%%%%%%%%%%%%%%%%%%%%%%%%%%%%%%%%%%%%%%%%%%%%%%%%%%%

\section{Variational Energy}
\label{sec:energy}

In the Rapid Communication, we present the extrapolated energies for different topological 
sectors and system sizes in the spin-liquid phase of the triangular Heisenberg model (THM). 
These are obtained from the \emph{variational} energy of 
well-converged wavefunctions using the $SU(2)$-symmetric infinite density-matrix 
renormalization-group (iDMRG) method with fixed number of states, $m$.
The extrapolation of the variational energy to
the thermodynamic limit of $m\rightarrow\infty$ needs particular care to achieve high accuracy.
During convergence of iDMRG sweeps, we notice that it is vital for 
the wavefunctions to be well-optimized with several sweeps of a
constant number of states, otherwise the resulting wavefunction may 
have an artificially higher variational energy leading to incorrect scaling.
The extrapolation to $m\rightarrow\infty$ can be done in several ways, here we compare
two well-known candidates, the \emph{energy variance} and DMRG \emph{truncation error}.

%%%%%%%%%%%%%%%%%%%%%%%%%%%%%%%%%%%%%%%%%%%%%%%%%%%%%%%%%%%%%%%%%%%%%%%%%%
\begin{figure}
  \begin{center}
    \includegraphics[width=0.64\linewidth]{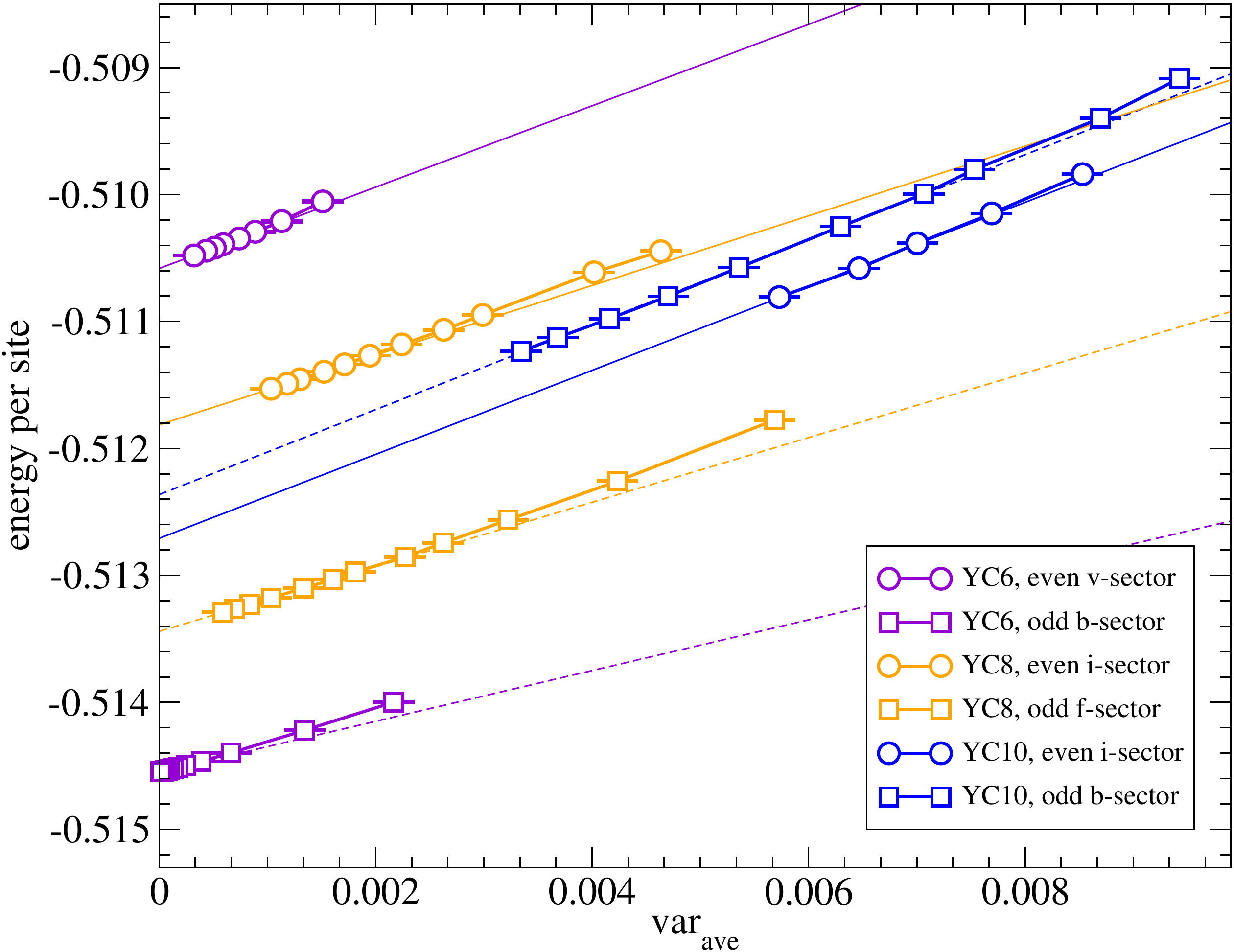}
    \caption{(Color online) 
    iDMRG results for the variational energy versus variance per site
    for the YC structures of the THM at $J_2=0.125$ (deep in 
    the $Z_2$ spin-liquid phase). The individual error bars are an estimate for iDMRG 
    systematic errors. Thin and dashed lines
    are linear fits to the final few points at largest basis sizes $m$, which
    gives the estimate of the energy in the thermodynamic limit (see below).
    \label{fig:energy-var}}
  \end{center}
\end{figure}
%%%%%%%%%%%%%%%%%%%%%%%%%%%%%%%%%%%%%%%%%%%%%%%%%%%%%%%%%%%%%%%%%%%%%%%%%%

The variance of the energy, $var_m = \langle \psi_m |(H-E_m)^2 | \psi_m \rangle$, is computationally 
somewhat costly to calculate for 2D cylinders due to the large dimension of the relevant operator 
in matrix product operator (MPO) form\cite{MPO}. However it is one 
of the strongest criteria for checking convergence of the wavefunctions, 
$| \psi_m \rangle$.  The variance \emph{per site} is well-defined in the thermodynamic
limit for iDMRG, and can be evaluated efficiently using the recursive approach described in \cite{Variance}. 
The results for variational energy versus variance per site 
(averaged over the $L_y$-site unit-cell), $var_{ave}$, is presented in \fref{fig:energy-var}. 
The individual error bars are an estimated DMRG systematic error when calculating the energy
for fixed basis size $m$, which we obtain from the minimum and maximum energies across a sweep
of the unit cell.

In \fref{fig:energy-var}, it is clear that the behavior of the variational energy is already indicative of
the general behavior (main letter, Fig.\ 2), even without the extrapolation to the thermodynamic 
limit of $m\rightarrow\infty$. The energy difference between even-boundary and odd-boundary topological 
sectors is rapidly decreasing with increasing $L_y$, which is consistent with the existence of a truly degenerate 
ground-state in the 2D limit of $L_y\rightarrow\infty$. We note that for narrower YC structures 
($L_y=6,8$), odd sector always has lowest energy, while for $L_y=10$, the even sector has the lowest energy. 
This crossover in the lowest energy state between even and odd sectors is not an artifact of the
extrapolation, since the variational energies already show the crossover behavior.

%%%%%%%%%%%%%%%%%%%%%%%%%%%%%%%%%%%%%%%%%%%%%%%%%%%%%%%%%%%%%%%%%%%%%%%%%%
\begin{figure}
  \begin{center}
    \includegraphics[width=0.64\linewidth]{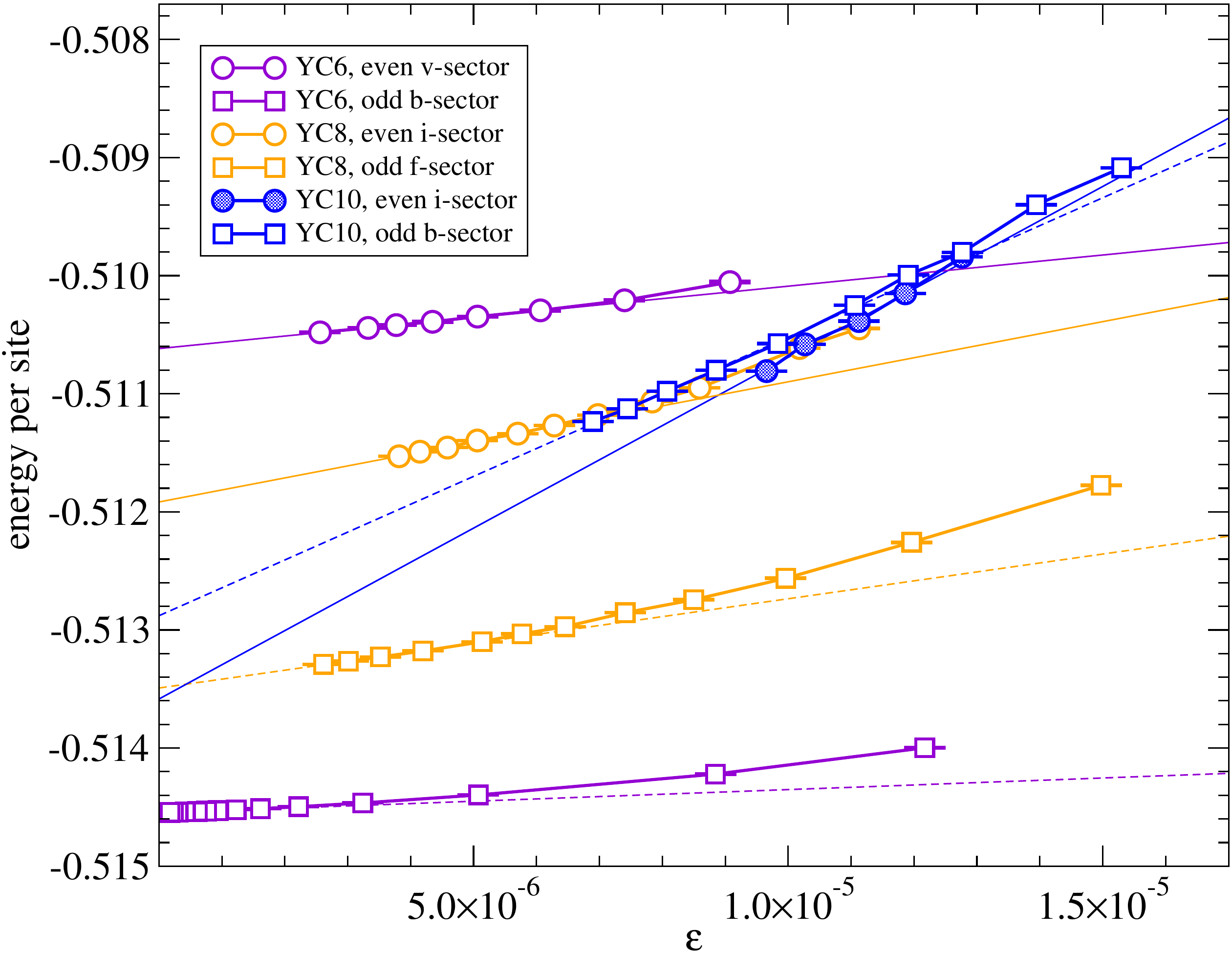}
    \caption{(Color online) 
    iDMRG results for the variational energy 
    versus truncation error for the YC structures of the THM at $J_2=0.125$.  
    The individual error bars are an estimate for iDMRG 
    systematic errors. Thin and dashed lines
    are linear fits to the final few points at largest basis sizes $m$, which
    gives the estimate of the energy in the thermodynamic limit (see below).
    \label{fig:energy-TruncError}}
  \end{center}
\end{figure}
%%%%%%%%%%%%%%%%%%%%%%%%%%%%%%%%%%%%%%%%%%%%%%%%%%%%%%%%%%%%%%%%%%%%%%%%%%

In DMRG, each step involves a truncation of the spectrum of the \textit{density matrix}, $\rho$.
This will produce a truncation error that is the sum of all discarded eigenvalues of $\rho$.
The actual value of this truncation error depends on the details of the algorithm -- for a purely
single-site algorithm with no density matrix mixing the truncation error is exactly zero, however
for two-site optimization and single-site algorithms with a non-zero density matrix mixing\cite{Hubig15},
the truncation error gives a stable quantity through which a scaling to the thermodynamic limit
$m \rightarrow \infty$ can be performed. Here, the truncation error, $\varepsilon_m$, of a 
state with a fixed $m$ means the average of all individual truncation
errors at each site of the unit cell over the final DMRG sweep. $\varepsilon_m$ is 
commonly considered as a good convergence criterion for the DMRG wavefunctions.
For well-converged wavefunctions in a DMRG calculation, 
the scaling behavior of energy with the truncation error is expected to be linear, while other
observables, $ \la \psi | O | \psi \ra$, generally scale with 
$\sqrt \varepsilon$\cite{White07}. However, we emphasize that the truncation error is highly algorithm-dependent,
and a small truncation error doesn't imply that the wavefunction is well converged. The
variance is a much more robust error estimator, and it is better to use this wherever possible.

We present the variational energy versus truncation error
in \fref{fig:energy-TruncError}. There are only minor differences with \fref{fig:energy-var}. 
It is worth mentioning that the small $m$-value wavefunctions usually cannot produce the correct scaling behavior, 
instead they may lead to some strong \emph{quadratic} behavior. 
The thermodynamic limit predictions of energies using truncation error fits matches closely to fitted 
results using variances, showing the high stability and convergence 
of the wavefunctions. However the difference is relatively large for YC10-\sector{i} (see below).  

We repeatedly observed an interesting exception to the smooth energy behavior shown in \fref{fig:energy-var} 
and \fref{fig:energy-TruncError}, for the even-sectors of the YC structure. 
There is a sudden drop in energy at a specific number of states, namely $m_{kink}$,
which depends strongly on the system size (also observed by Zu \emph{et al.}\cite{Zhu-White_DMRG}). 
This occurs for a number of states $m_{kink} \sim 10^3$ for $L_y\geq10$. 
Initially when $m < m_{kink}$, the energy is in a higher meta-stable plateau (not shown in the figures), 
even for apparently well-converged wavefunctions. But the energy will suddenly drop to a lower plateau for 
$m > m_{kink}$, and thereafter show a systematic linear extrapolation with variance as expected. 
This is a common phenomena in DMRG calculations on systems with inhomogeneous 
ground-states. Our tests show that the wavefunctions before the $m_{kink}$ lack 
important geometrical symmetry properties (e.g.\ reflection $\la R_y \ra$) 
and give an incorrect scaling for the energy.
As a result, we disregard any wavefunction with $m < m_{kink}$ in the results. 
However, such states may preserve some subset of Hamiltonian symmetries. 
In particular, we found that translation symmetry in the $Y$-direction is well-preserved, 
which leads to a surprisingly well-converged appearance of the `momentum-resolved' entanglement spectrum (ES)
(see below), which is a pitfall for DMRG calculations -- one must perform careful checking to ensure that \emph{all}
expected symmetries are preserved.

Hu \emph{et al.}\cite{Hu-Sheng_DMRG} also present bulk ground-state energies versus truncation errors 
for fixed-$m$ wavefunctions in both even- and odd-boundary topological sectors for YC10, using DMRG on 
\emph{finite}-width cylinders. For extrapolating individual energies, they employ 
a \emph{quadratic} fit, presumably due to larger quadratic corrections from
calculating energies in the bulk of the system (excluding the
sites near the boundary). Nevertheless, finite DMRG and iDMRG individual energies 
are in a relatively good match, especially for the even-boundary topological sector.

%%%%%%%%%%%%%%%%%%%%%%%%%%%%%%%%%%%%%%%%%%%%%%%%%%%%%%%%%%%%%%%%%%%%%%%%%%%
\begin{figure}
  \begin{center}
    \includegraphics[width=0.44\columnwidth]{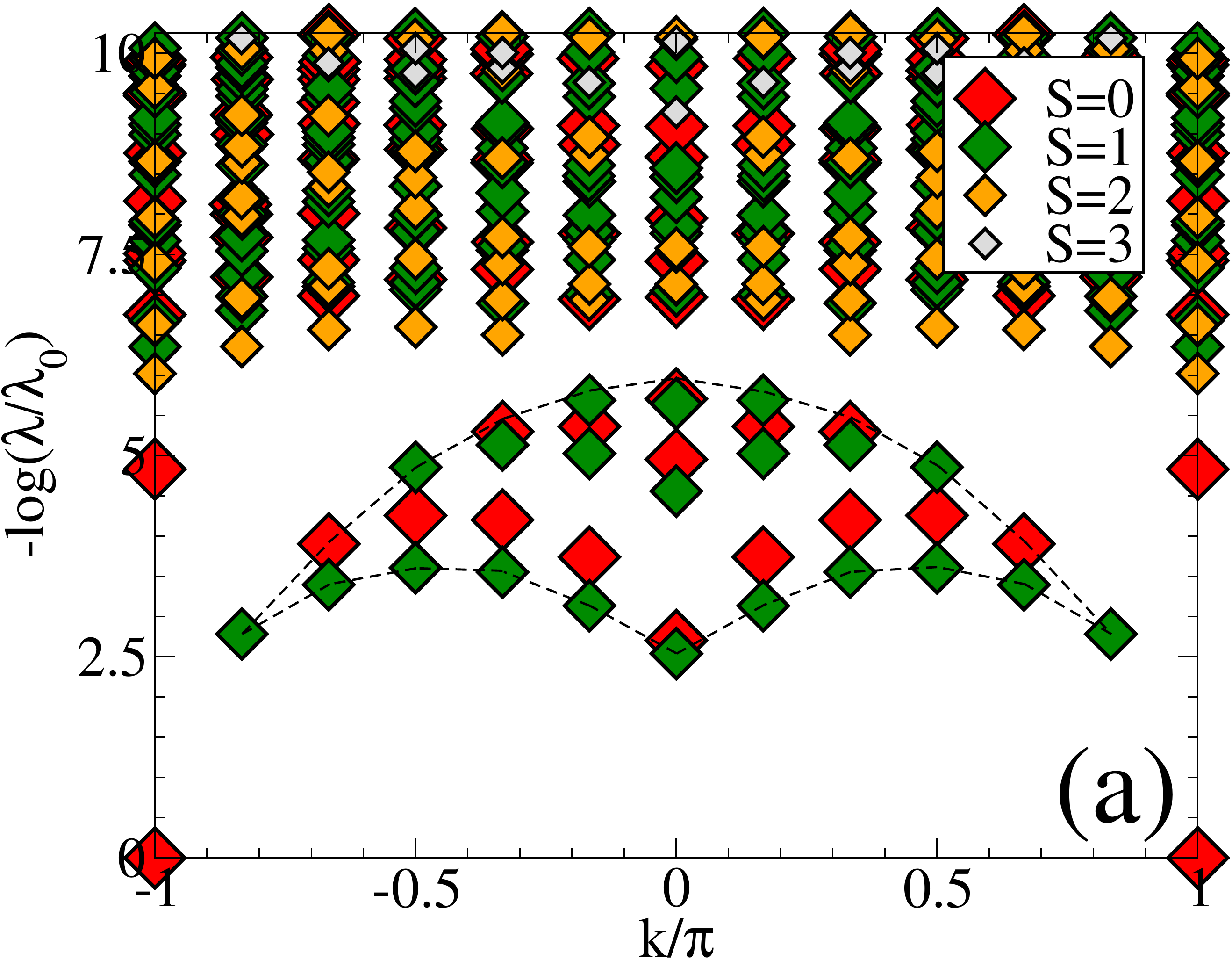}
    \includegraphics[width=0.44\columnwidth]{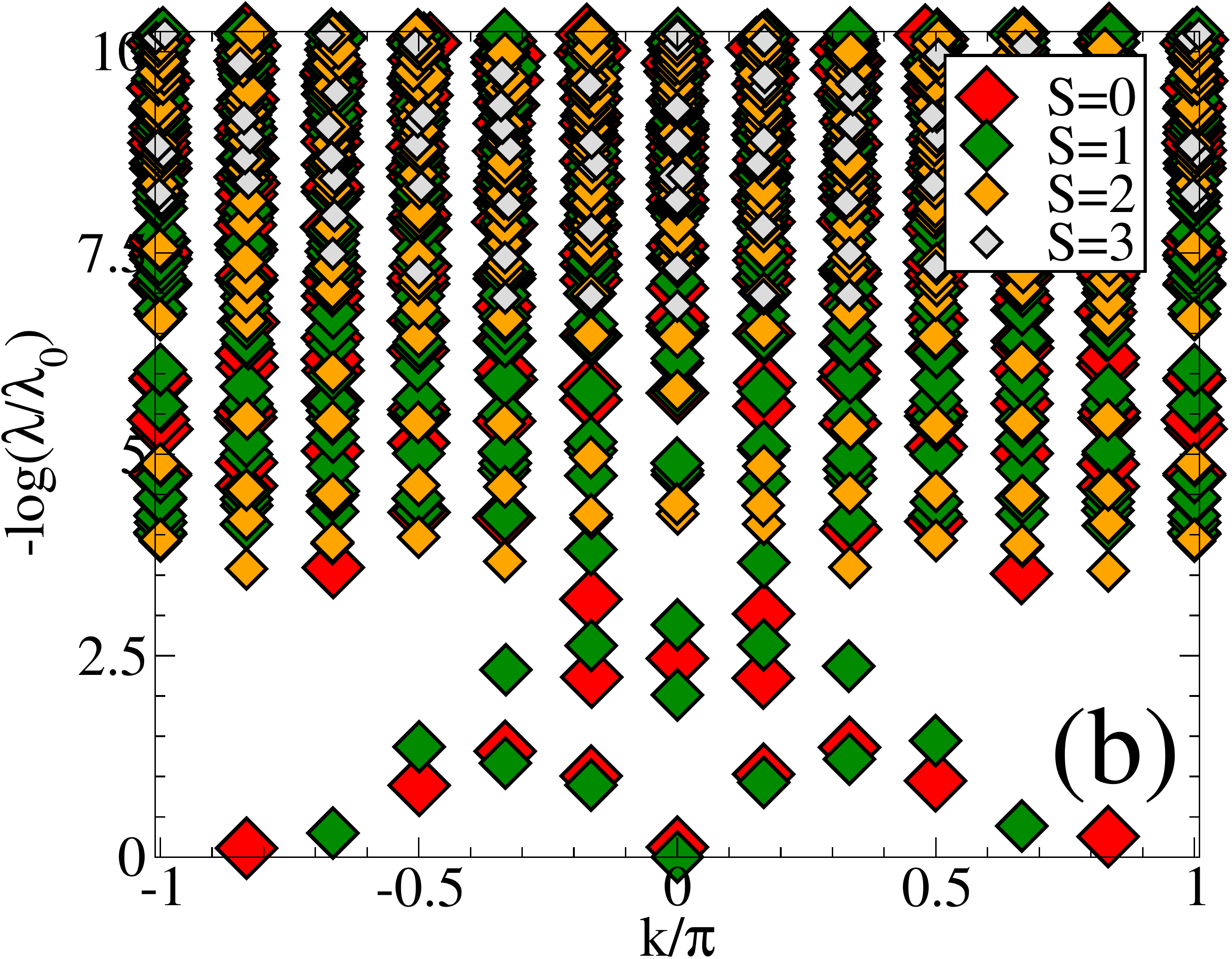}
    \caption{(Color online) 
    The evolution of ES in the even sectors while
    increasing the number of states, $m$. (a) A wavefunction 
    with $m \lesssim m_{kink}$. Dashed lines only connect the boundaries of 
    low-lying states to emphasize a candidate for the two-spinon excitation 
    spectrum. (b) A better-converged ground-state
    with $m \gtrsim m_{kink}$, consistent with stabilization of an 
    \sector{i}-sector. Spectra belong to iDMRG results for
    YC12 structures of the THM at $J_2=0.125$. 
    \label{fig:ES-YC12}}
  \end{center}
\end{figure}
%%%%%%%%%%%%%%%%%%%%%%%%%%%%%%%%%%%%%%%%%%%%%%%%%%%%%%%%%%%%%%%%%%%%%%%%%%%

We also performed iDMRG calculations for the even- and odd-sectors of the YC12 structure. 
Our individual wavefunctions for YC12 are not as well converged as the narrower cylinders,
even up to $m=5000$ $SU(2)$ states, so we have not included the energy extrapolations in our main results.
In particular, it appears that most of YC12 even sector results are still below 
$m_{kink}(L_y=12)$. However we were able to produce few points slightly above 
$m\approx5000$, where a \emph{valid} ES was obtained, and consistent with the results for smaller systems
(main Letter, Fig.~3). We show this spectrum, as well as another spectrum \emph{above} the kink, for
YC12 in \fref{fig:ES-YC12}. The ES below the kink, \fref{fig:ES-YC12}(a), 
which we emphasize does not represent the ground-state,
nevertheless has a low-lying structure that is in one-to-one correspondence with a finite-size two-spinon
spectrum\cite{SpinonSpectrum_theory}. This feature appears to be generic to weakly-coupled rings
(also seen in well-converged wavefunctions by reducing the magnitude of the couplings in the $X$-direction).
The ES for the ground-state, \fref{fig:ES-YC12}(b), appears to 
be a heavily distorted version of the two-spinon continuum, probably
with a \emph{quadratic} rather than linear dispersion around $k=\pm\pi$.

\section{Detailed results for symmetry group measurements}
\label{sec:DetailedList}

%%%%%%%%%%%%%%%%%%%%%%%%%%%%%%%%%%%%%%%%%%%%%%%%%%%%%%%%%%%%%%%%%%%%%%%%%%%%%%%%%%%%%%%%%%%%%%%%%%%%%%%%%%%%%%%%%%%%%%%%%%%%%%%%%%%%%%%%
%\newcommand\T{\rule{0pt}{2.6ex}}
\begin{table*}
  \caption{Detailed list of energy and symmetry group measurements for the THM at $J_2=0.125$. 
  $\langle R_y \rangle$ is calculated per enlarged unit cell of $L_y \times L_y$ sites, which partially
  accounts for the reduced accuracy of this quantity compared with the other results. The $Y$-momentum, $\theta_y$,
  calculated from the eigenvalue of the translation operator, $T_y$, is obtained per unit cell of $L_y$ sites,
  which is numerically indistinguishable from $\pm 1.0$.
  \label{table:DetailedList}}
  \begin{center}
   %\begin{ruledtabular}
    \begin{tabular}{ |c|c|c|c|r@{$ 1 \; \pm \; $}l|r@{$ 0. $}l|c|r@{$ 0. $}l|r@{$ 0. $}l|r@{$ 1 \; \pm \; $}l|c| }
        \hline \hline
        \tiny Structure \normalsize & \tiny \specialcell[c]{spin-$\frac{1}{2}$\\boundary} \normalsize \T & \scriptsize \specialcell[c]{
        ES\\degeneracy} \normalsize & \scriptsize \specialcell[c]{extrapolated\\energy per site} \normalsize & \multicolumn{2}{c|}{$\la 
        C[P^2] \ra$} & \multicolumn{2}{c|}{$\la R_y \ra$} & $\theta_y$ & \multicolumn{2}{c|}{$\la C[D_{L_y}] \ra$} & 
        \multicolumn{2}{c|}{$\la C[T_y^\pi,\tau] \ra$} & \multicolumn{2}{c|}{$\la C[\tau^2] \ra$} & \tiny \specialcell[c]{
        sector} \normalsize \\ \hline
        %%%%%%%%%%%%%%%%%%%%%%%%%%%%%%%%%%%%%%%%%%%%%%%%%%%%%%%%%%%%%%%%%%%%%%%%%%%%%%%%%%%%%%%%%%%%%%%%%%%%%%%%%%%%%%%%%%%
        YC6 & even & 2-fold & \scriptsize \specialcell[c]{$-0.510617(6)_{Trunc}$\\$-0.510582(2)_{Var}$} 
        \normalsize & & $10^{-7}$ & & $9998$ & $\pi$ & $-$ & $999996$ & $-$ & $999994$ & & $10^{-11}$ \T & \sector{v} \\ 
        \hline
        %%%%%%%%%%%%%%%%%%%%%%%%%%%%%%%%%%%%%%%%%%%%%%%%%%%%%%%%%%%%%%%%%%%%%%%%%%%%%%%%%%%%%%%%%%%%%%%%%%%%%%%%%%%%%%%%%%%
        YC6 & odd & 2-fold & \scriptsize \specialcell[c]{$-0.5145491(2)_{Trunc}$\\$-0.5145485(2)_{Var}$} 
        \normalsize & $-$ & $10^{-8}$ & & $999993$ & $0$ & & $9999998$ & $-$ & $9999998$ & $-$ & $10^{-14}$ \T & \sector{b} \\ 
        \hline
        %%%%%%%%%%%%%%%%%%%%%%%%%%%%%%%%%%%%%%%%%%%%%%%%%%%%%%%%%%%%%%%%%%%%%%%%%%%%%%%%%%%%%%%%%%%%%%%%%%%%%%%%%%%%%%%%%%%
        YC8 & even & \scriptsize \specialcell[c]{non-\\degenerate} \normalsize & \scriptsize \specialcell[c]{
        $-0.51192(2)_{Trunc}$\\$-0.511814(6)_{Var}$} \normalsize & & $10^{-6}$ & & $998$ & $0$ & & $99998$ & & $99997$ & 
        & $10^{-10}$ \T & \sector{i} \\ \hline
        %%%%%%%%%%%%%%%%%%%%%%%%%%%%%%%%%%%%%%%%%%%%%%%%%%%%%%%%%%%%%%%%%%%%%%%%%%%%%%%%%%%%%%%%%%%%%%%%%%%%%%%%%%%%%%%%%%%
        YC8 & odd & 4-fold & \scriptsize \specialcell[c]{$-0.513492(7)_{Trunc}$\\$-0.513441(1)_{Var}$} \normalsize & $-$ & 
        $10^{-7}$ & $-$ & $9994$ & $\pi$ & $-$ & $999990$ & & $99998$ & $-$ & $10^{-11}$ \T & \sector{f} \\ \hline
        %%%%%%%%%%%%%%%%%%%%%%%%%%%%%%%%%%%%%%%%%%%%%%%%%%%%%%%%%%%%%%%%%%%%%%%%%%%%%%%%%%%%%%%%%%%%%%%%%%%%%%%%%%%%%%%%%%%
        YC10 & even & \scriptsize \specialcell[c]{non-\\degenerate} \normalsize & \scriptsize \specialcell[c]{
        $-0.5136(2)_{Trunc}$\\$-0.5127(1)_{Var}$} \normalsize & & $10^{-7}$ & $-$ & $98$ & $\pi$ & & $9996$ & & $9996$ & 
        & $10^{-9}$ \T & \sector{i} \\ \hline
        %%%%%%%%%%%%%%%%%%%%%%%%%%%%%%%%%%%%%%%%%%%%%%%%%%%%%%%%%%%%%%%%%%%%%%%%%%%%%%%%%%%%%%%%%%%%%%%%%%%%%%%%%%%%%%%%%%%
        YC10 & odd & 2-fold & \scriptsize \specialcell[c]{$-0.51288(4)_{Trunc}$\\$-0.51236(1)_{Var}$} \normalsize & $-$ & 
        $10^{-6}$ & & $993$ & $0$ & & $9998$ & $-$ & $9998$ & $-$ & $10^{-9}$ \T & \sector{b} \\  
        %%%%%%%%%%%%%%%%%%%%%%%%%%%%%%%%%%%%%%%%%%%%%%%%%%%%%%%%%%%%%%%%%%%%%%%%%%%%%%%%%%%%%%%%%%%%%%%%%%%%%%%%%%%%%%%%%%%
        \hline \hline
    \end{tabular} 
   %\end{ruledtabular}
  \end{center}
\end{table*}
%%%%%%%%%%%%%%%%%%%%%%%%%%%%%%%%%%%%%%%%%%%%%%%%%%%%%%%%%%%%%%%%%%%%%%%%%%%%%%%%%%%%%%%%%%%%%%%%%%%%%%%%%%%%%%%%%%%%%%%%%%%%%%%%%%%%%%%%

In the main letter, we understand that the obtained structure of four anyon sectors identifies the symmetry as 
corresponding to the ordering of an $Z_2$ RVB with anyonic quasi-particles. The fusion rules\cite{Kitaev06} of these
anyons read: $\sector{b}\times \sector{i}=\sector{b}$, 
$\sector{v}\times\sector{i}=\sector{v}$, $\sector{f}\times\sector{i}=\sector{f}$, $\sector{b}\times\sector{v}=\sector{f}$, and 
$\sector{i}\times\sector{i}=\sector{v}\times\sector{v}=\sector{b}\times\sector{b}=\sector{f}\times\sector{f}=\sector{i}$ 
(anyons are their own anti-particle). Each anyonic sector corresponds to 
a particle/anti-particle pair located at each end of the infinite cylinder,
which is exactly the MES state. It is well-known\cite{Jiang12,Zaletel-PRL} 
that MESs are local minima of entanglement entropy, so we expect iDMRG 
to naturally converge toward them, although stabilizing all anyonic sectors proved to be
numerically challenging. We then present an analysis of time-reversal symmetry, $\tau$,
which can identify the existence of spinon excitations\cite{SpinonExcitation}  
and projective symmetry groups (PSGs) of the cylinder's dihedral group, $D_{L_y}$, which can 
identify the existence of vison excitations\cite{VisonExcitation} (already classifying all sectors).
We note that the sign of proposed commutators, $\la C[D_{L_y}] \ra$ and $\la C[\tau^2] \ra$, is \emph{not} fixed for the
$\sector{v}$-sector due to criteria of `generalized flux-fusion anomaly test', as proposed by 
Qi \emph{et~al.}\cite{Qi15} for different topological invariants when explicitly preserving SU(2) symmetry. 
In \tref{table:DetailedList}, we present additional symmetry and energy measurements 
for the different topological sectors and system sizes. For comparison purposes, we first report extrapolated 
energies using truncation error fits ($E_{trunc}$) and variance 
fits ($E_{var}$) from \fref{fig:energy-TruncError} and \fref{fig:energy-var}
respectively. Parity reflection ($P$) here is a reversal of the ordering of the MPS wavefunction,
which corresponds to both a reflection in the $X$-direction (exchanging boundaries of the cylinder), as well
as a reversal of ordering of sites within the unit cell, which corresponds to a reflection in the $Y$-direction.
Similarly to the case of time reversal, the parity reflection operator functions as an anti-linear operator acting on the
entanglement Hamiltonian so we obtain the square of parity reflection measured on the MPS auxiliary basis, 
$ C[P^2] = \la \Upsilon_P \Upsilon_P^* \ra$, in a similar way to $C[\tau^2]$. The 
results also follow exactly $C[\tau^2]$ as clear from \tref{table:DetailedList}. I.e.\ the SPT 
structure of \sector{b} and \sector{f}-sectors are simultaneously 
protected by parity reflection and time-reversal symmetries.
 
Next, we present the eigenvalues of reflection in the $Y$ direction, $R_y$. There are 
subtleties involved in this measurement. Because of the form of the wrapping of the cylinder Hamiltonian
onto a one-dimensional MPS, the central site (or bond) that the reflection is applied
around is changing when one moves from one unit cell of the cylinder to the next. To compensate for this,
the final form of $R_y$ needs to be written as an operator acting on a unit cell of size $L_y \times L_y$,
which is a rather complicated object. As a result, $\langle R_y \rangle$ is obtained as an eigenvalue measured
per $L_y \times L_y$ sites, thus particularly sensitive to convergence of the wavefunction.
We find that the closeness of $|\la R_y \ra|$ to the identity is the hardest
convergence criterion for the variational wavefunctions to meet. The wavefunctions
presented in \tref{table:DetailedList} achieve the best $|\la R_y \ra|$ within our 
available computational resources. 

By comparison, construction of translation in the $Y$-direction, $T_y$, is relatively straightforward
as a product of swap operations acting purely within one unit cell, hence $\la T_y \ra$ is obtained as
an eigenvalue per $L_y$ sites. Since convergence of $\la T_y \ra$ is easily
achievable for variational wavefunctions (even keeping a relatively small basis size $m$), we only 
report the complex phase of $\la T_y \ra$, i.e.\ the ground-state momentum 
in the $Y$-direction ($\theta_y$) in the table. $\theta_y$ is consistent
with the ground-state momentum of a spin-$\frac{1}{2}$, $L_y$-site Heisenberg ring\cite{OriginalMomentum} for the 
even sectors, and the odd sectors have a relative momentum shift of $\pi$. 

We also report results of measurement of another observable, 
$C[T_y^\pi,\tau] = (T_y^\pi \tau)^2 . (\tau)^2 = T_y^\pi \tau T_y^\pi \tau^{-1}$. In auxiliary
basis notation, one can write $\la C[T_y^\pi,\tau]\ra = \langle \Upsilon_{T_y^\pi} 
\Upsilon_\tau \Upsilon_{T_y^\pi}^* \Upsilon_{\tau}^\dagger \rangle$, which denotes the anti-symmetry of
the combination of time-reversal and a $\pi$ rotation around the $Y$-axis and 
is not an independent topological invariant, but the product of $C[D_{L_y}]$ and $C[\tau^2]$.

\section{Entanglement Entropy}
\label{sec:EE}

%%%%%%%%%%%%%%%%%%%%%%%%%%%%%%%%%%%%%%%%%%%%%%%%%%%%%%%%%%%%%%%%%%%%%%%%%%%
\begin{figure}
  \begin{center}
    \includegraphics[width=0.44\linewidth]{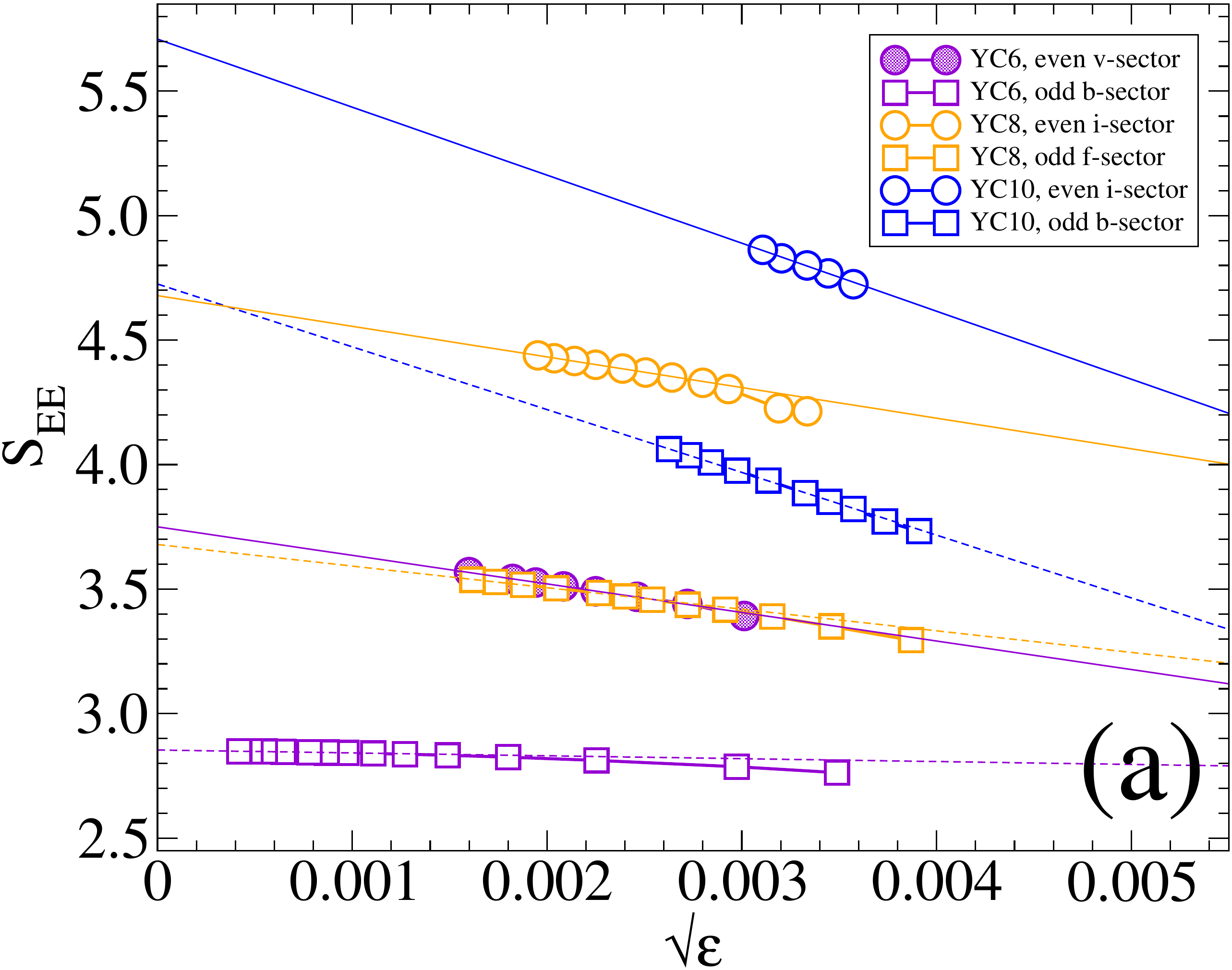}
    \includegraphics[width=0.44\linewidth]{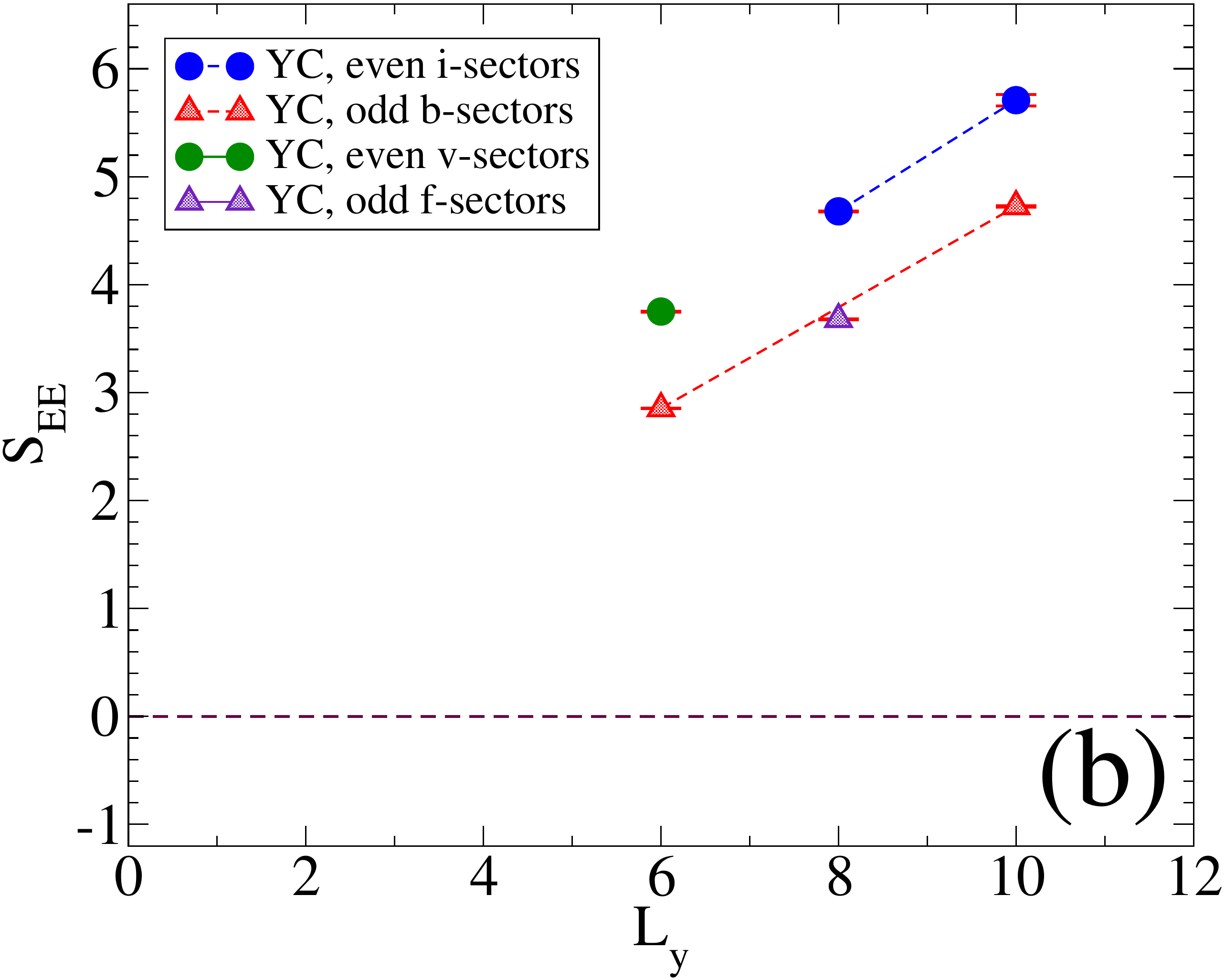}
    \caption{(Color online) 
    iDMRG results for the EE in the YC structures of the THM at $J_2=0.125$. 
    (a) Individual entropies for different topological sectors and 
    system sizes against square root of truncation errors, 
    $\sqrt{\varepsilon}$. Thin and dashed lines are linear fits for the points with
    smallest truncation errors.
    (b) Extrapolated entropies at the thermodynamic limit of 
    $m\rightarrow\infty$, for different topological sectors and system sizes.
    \label{fig:entropy}}
  \end{center}
\end{figure}
%%%%%%%%%%%%%%%%%%%%%%%%%%%%%%%%%%%%%%%%%%%%%%%%%%%%%%%%%%%%%%%%%%%%%%%%%%%  

The entanglement entropy\cite{Eisert10} (EE) is a central object in the fields of 
quantum information and many-body physics. It is widely used to identify the nature of the low-energy
spectrum, symmetry breaking, and topological degeneracy of the ground-state (see for example 
Refs.~\onlinecite{Kitaev-Preskill,Levin06,Eisert10,Zhang11,Wang13,Metlitski15}). The EE can be measured in different ways, but 
the calculation of the von Neumann entropy along the bipartition cut of the cylinder\cite{Jiang12} 
is computationally easy from the MPS ansatz, as
$S_{EE} = - tr(\rho_r \log\rho_r)$, where $\rho_r$ is the reduced density 
matrix of either partition of the cut. 

For two-dimensional quantum systems with \emph{local} interactions and a boundary size significantly larger 
than the correlation length, it is well-known\cite{AreaLaw,Eisert10,Kitaev-Preskill,Levin06} that the EE will scale with 
the boundary area (so-called \emph{area-law}), not the system volume. 
In this case, one can write\cite{Kitaev-Preskill,Levin06} $S_{EE} = \alpha L_y - \gamma + ...$, where $\alpha$ is a 
non-universal constant depending on short wavelength modes near the boundary, $L_y$ is the size of the cut, 
$\gamma$ is the topological entanglement entropy (TEE) that depends only on the quantum dimension 
of the ground-state, and we ignore corrections that vanish in the thermodynamic limit.
The TEE depends on the ground-state degeneracy of the system, and is $\ln(2) \approx 0.6931$ 
for the $Z_2$ topological phase (see for example Refs.~\onlinecite{Zhang11,Jiang12}).    

We present our results for the EEs of the individual wavefunctions in \fref{fig:entropy}(a) and 
extrapolated values in \fref{fig:entropy}(b). As it is clear from \fref{fig:entropy}(a), 
for fixed-$L_y$ the even sector always has higher EE and support more entanglement than the odd sector 
for the same number of states. Reliable extrapolations to the thermodynamic 
limit of $m\rightarrow\infty$ are possible for both sectors by employing a linear fit against 
$\sqrt{\varepsilon}$ as shown in the figure. However, in \fref{fig:entropy}(b), no obvious fit for the EE 
in the 2D limit, $L_y \rightarrow \infty$, can be done that produces a non-spurious TEE close to 
$\gamma=\ln(2)$, most likely due to finite size effects.
The direct measurement of the TEE, using linear extrapolations to entropies derived for a 
fixed-$L_y$ cylinders, faces several obstacles. The main problem is that 
two of our three major system sizes, $L_y=6,8$, are quite likely still too narrow to produce 
correct scaling for extrapolation in large-$L_y$. 
In addition, $L_y=10$ is exactly where DMRG studies (cf. Ref.\ \onlinecite{Hu-Sheng_DMRG} and the 
main Letter) observe a crossing between energy of different
topological sectors. A similar effect was also seen by Iqbal \emph{et. al}\cite{Iqbal16}, using Quantum
Monte-Carlo, who found that the energy scaling starts to change considerably for $L_y\geq10$. 
This suggests that only system sizes of $L_y\geq10$ can effectively be 
employed to extrapolate observables for the 2D limit of $L_y\rightarrow\infty$. 
Additionally, we have obtained at most two different system sizes per topological sector.
In principle for finite width systems the EE values 
can lie on different lines for each sector, which makes it hard to trust in any 
entropy extrapolation containing different anyonic sectors. The existence of 
some \emph{gapless} modes in the spin-liquid phase at thermodynamic limit would give a $log$ 
correction\cite{Wang13} to the entropy, which would make a fit to a few data points prohibitively difficult.
As a result of these obstacles, we conclude that the set of $L_y=6,8,10$ are not expected to 
produce any reliable TEE value even having individually highly-converged wavefunctions.

\section{Bond Anisotropies}
\label{sec:BondAnisotropy}

In the main Letter we present the real-space correlations (lattice
visualization) of a YC10-\sector{b} system. The overall pattern of NN correlations (bonds)
is identical for all wavefunctions in odd-boundary 
topological sectors with $L_y=8,10$. In \fref{fig:LatticeVisualisation},
we present a lattice visualization for the YC10-\sector{i} system, as a representative 
picture for all even-boundary topological sectors, which demonstrates their general pattern
of bond anisotropies.

%%%%%%%%%%%%%%%%%%%%%%%%%%%%%%%%%%%%%%%%%%%%%%%%%%%%%%%%%%%%%%%%%%%%%%%%%%%
\begin{figure}
  \begin{center}
    \includegraphics[width=0.64\columnwidth]{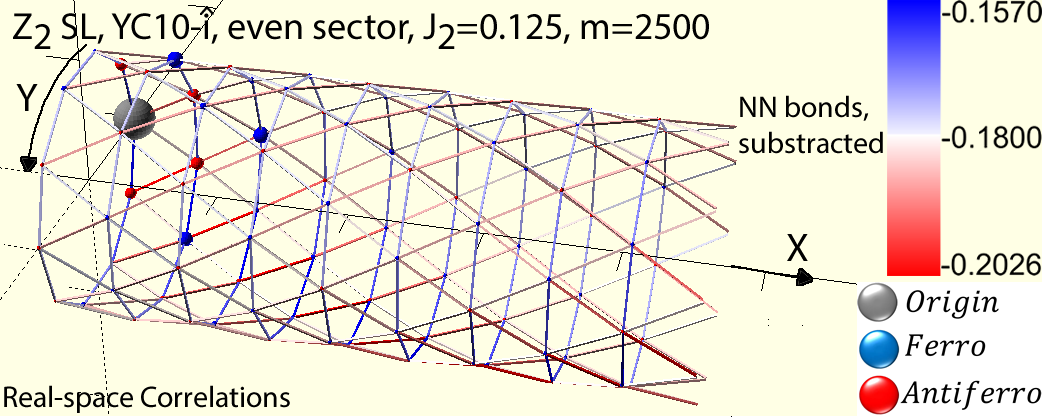}
    \caption{(Color online) 
    Lattice visualization of a YC10-\sector{i}
    wavefunction. NN and long-range correlations are calculated using iDMRG, then the
    size and the color of bonds and spin spheres are set accordingly. The average of NN 
    correlation is subtracted from each bond to highlight the anisotropy pattern.   
    \label{fig:LatticeVisualisation}}
  \end{center}
\end{figure}
%%%%%%%%%%%%%%%%%%%%%%%%%%%%%%%%%%%%%%%%%%%%%%%%%%%%%%%%%%%%%%%%%%%%%%%%%%% 

\begin{figure}
  \begin{center}
    \includegraphics[width=0.44\linewidth]{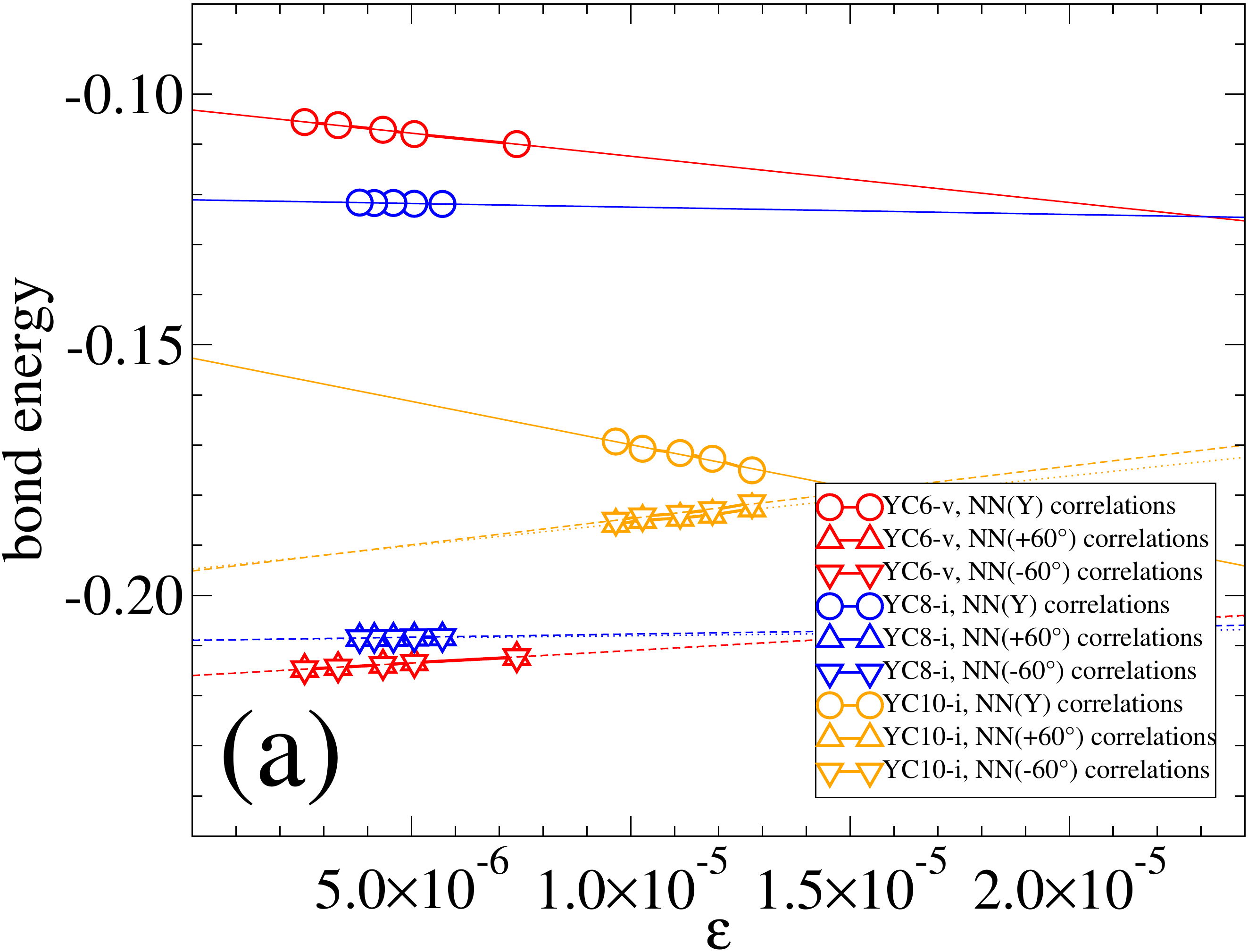}
    \includegraphics[width=0.44\linewidth]{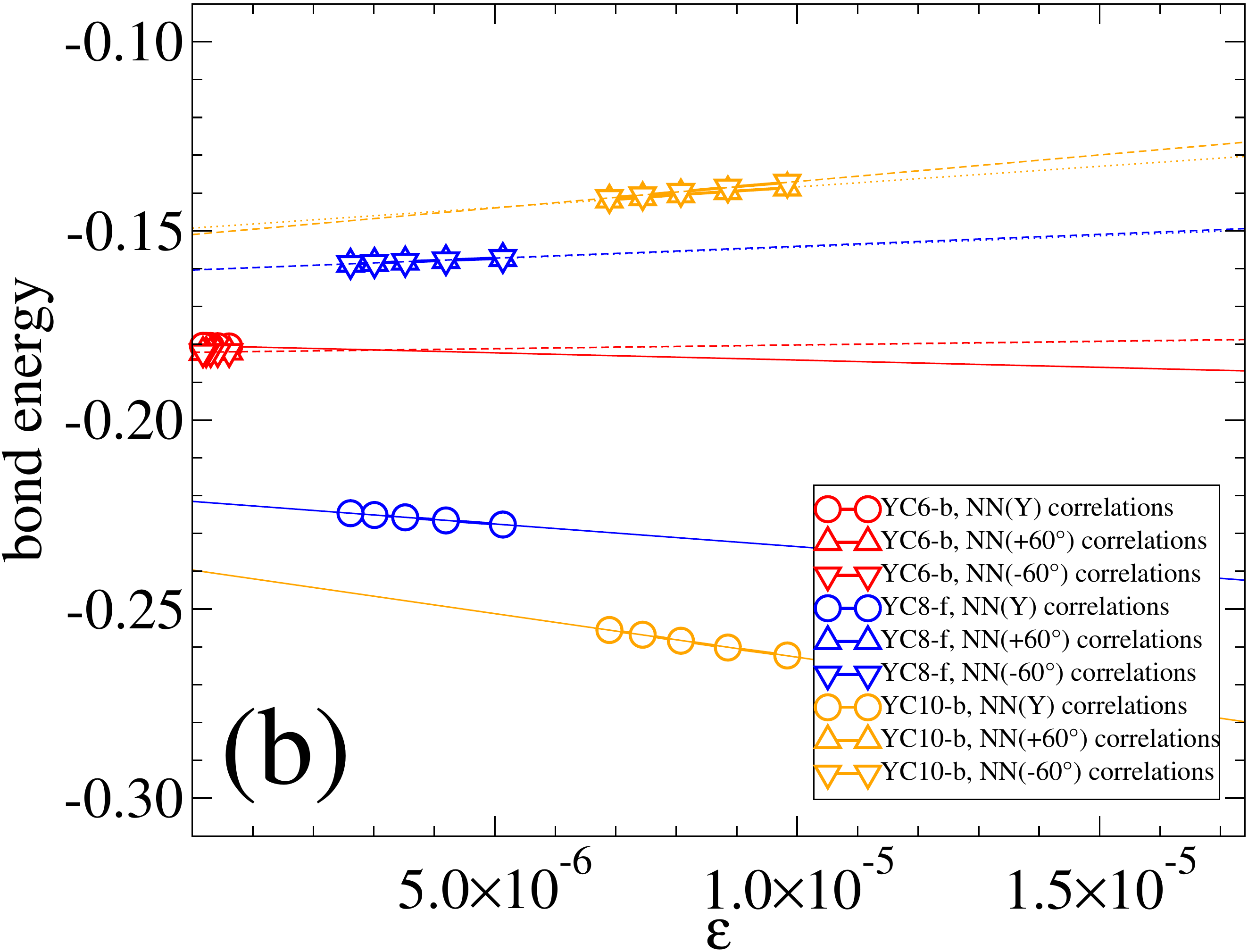}
    \caption{(Color online) 
    iDMRG results for the NN bond anisotropies of the THM in three lattice directions
    versus the truncation error, $\varepsilon$, at $J_2=0.125$. Each colored 
    series of results presents bond energies of a YC$L_y$ structure. (a) Even sectors. 
    (b) Odd sectors. Thin lines and dashed-lines are linear fits to 
    serve as the extrapolated NN bond energies in the thermodynamic 
    limit of $m\rightarrow\infty$ (see below).
    \label{fig:BondAnisotropy}}
  \end{center}
\end{figure}

We observe relatively strong patterns of bond anisotropy in 
YC structures of the spin liquid (SL). If the ground-state preserves all the 
continuous symmetries of the Hamiltonian and the point group symmetries of the lattice, 
one expects to observe \emph{no} bond anisotropy pattern in the 2D limit of 
$L_y\rightarrow\infty$. If there is persistent existence of bond anisotropies in the lattice 
principal directions, the phase can be interpreted as a 
``nematic'' SL\cite{NematicSL}, which spontaneously breaks $\frac{2\pi}{3}$ 
rotational symmetry (i.e. ${\bf C}_6$ group) of the triangular lattice,
while preserving mirror symmetries. To provide more clarification on the nature of 
the bond anisotropies in the large system size limit, we calculate the NN 
correlations, averaged over a unit-cell, in all three principal directions of the lattice. Results for
the bond anisotropies as a function of truncation error are presented
in \fref{fig:BondAnisotropy}(a) for the even sectors and in \fref{fig:BondAnisotropy}(b) 
for the odd sectors.  

%%%%%%%%%%%%%%%%%%%%%%%%%%%%%%%%%%%%%%%%%%%%%%%%%%%%%%%%%%%%%%%%%%%%%%%%%%%
%%%%%%%%%%%%%%%%%%%%%%%%%%%%%%%%%%%%%%%%%%%%%%%%%%%%%%%%%%%%%%%%%%%%%%%%%%%

\begin{figure}
  \begin{center}
    \includegraphics[width=0.64\linewidth]{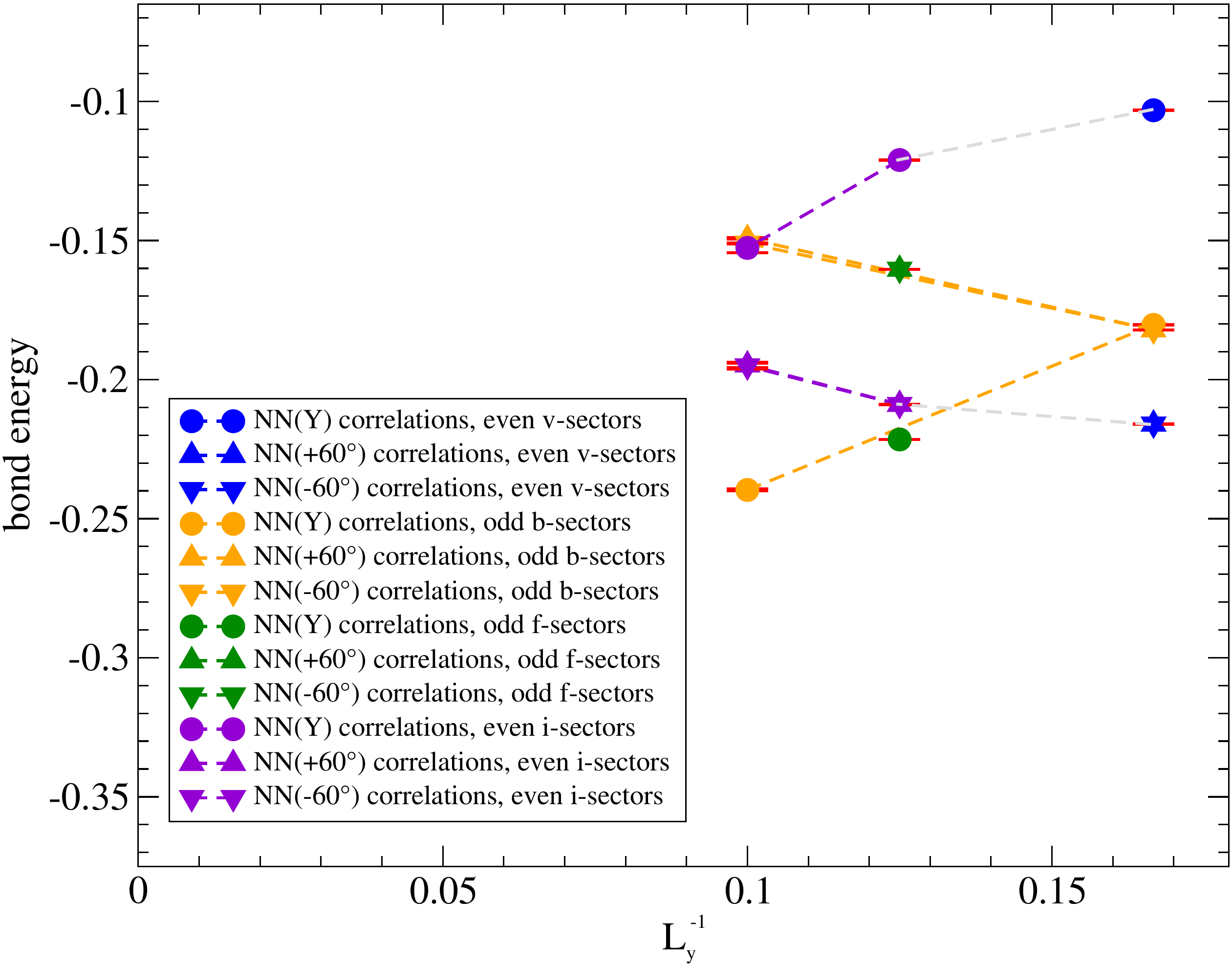}
    \caption{(Color online) 
    iDMRG results for the extrapolated NN bond anisotropies of the THM 
    in the three lattice directions against inverse of the system width. 
    Results belong to topological sectors with YC structures at $J_2=0.125$. 
    The individual error-bars are the standard deviations of linear fits 
    against truncation errors, presented in \fref{fig:BondAnisotropy}(a) 
    and \fref{fig:BondAnisotropy}(b) .
    \label{fig:BondAnisotropy-extrapolated}}
  \end{center}
\end{figure}

We present extrapolated NN bond energy results (for the thermodynamic limit of 
$m\rightarrow\infty$) in \fref{fig:BondAnisotropy-extrapolated}. Evidently, even-boundary sectors
\sector{v} and \sector{i} are becoming more symmetric (less anisotropic) while 
increasing $L_y$, and it is reasonable to suppose that these sectors become isotropic in the 2D limit.
However, for the odd-boundary \sector{b} and \sector{f} sectors, the anisotropy is increasing
with increasing $L_y$. This is more consistent with the existence 
of a nematic SL phase in the 2D limit. However odd-sector results may be affected more strongly by 
the finite size effects of $L_y$ due to the placement of a half-integer quantum-number 
on the unit cell boundary. Similar patterns of the bond anisotropies are 
observed for both even- and odd-sectors by Hu \emph{et al.}\cite{Hu-Sheng_DMRG}. 

%%%%%%%%%%%%%%%%%%%%%%%%%%%%%%%%%%%%%%%%%%%%%%%%%%%%%%%%%%%%%%%%%%%%%%%%%%%
%%%%%%%%%%%%%%%%%%%%%%%%%%%%%%%%%%%%%%%%%%%%%%%%%%%%%%%%%%%%%%%%%%%%%%%%%%%

\section{Correlation Lengths}
\label{sec:CorrLength}

%%%%%%%%%%%%%%%%%%%%%%%%%%%%%%%%%%%%%%%%%%%%%%%%%%%%%%%%%%%%%%%%%%%%%%%%%%%
\begin{figure}
  \begin{center}
    \includegraphics[width=0.64\linewidth]{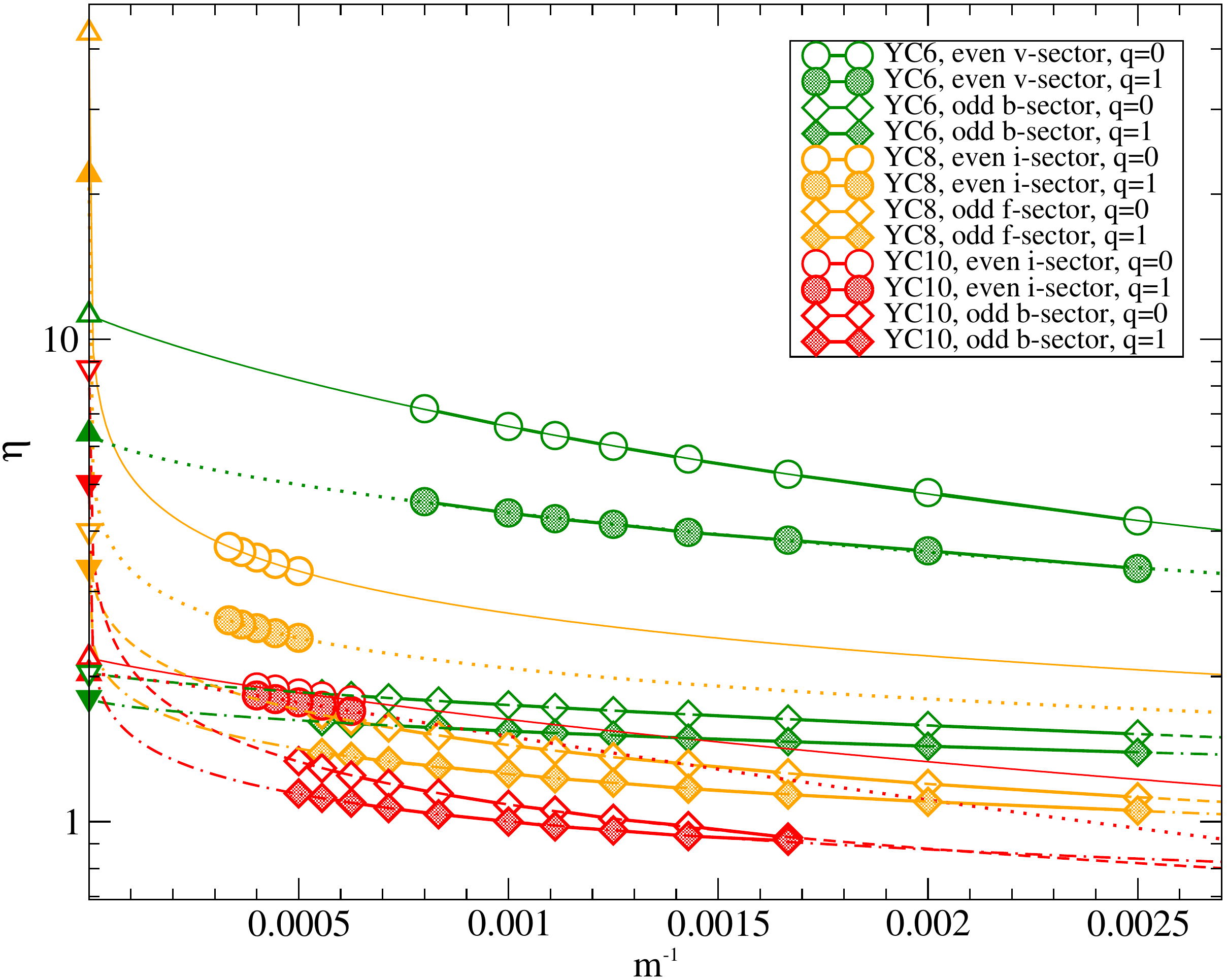}
    \caption{(Color online) 
    iDMRG results for the correlation length per unit-cell size 
    against inverse of the number of states, $m$, in different anyonic sectors and 
    system sizes of THM at $J_2=0.125$. Results are labeled with SU(2) 
    quantum numbers of $q=0,1$. Thin and dashed lines are
    attempted power-law fits to data according to $y^{-1} = a_0 + a_1 x^{a_2}$.
    Triangular symbols show extrapolated values of $a_0$ in the
    $m\rightarrow\infty$ limit.
    \label{fig:CorrLength}}
  \end{center}
\end{figure}
%%%%%%%%%%%%%%%%%%%%%%%%%%%%%%%%%%%%%%%%%%%%%%%%%%%%%%%%%%%%%%%%%%%%%%%%%%%

The nature of the \emph{spin gap} is a key property of the system. Static correlation functions are
closely related to the inverse gap size for both finite-width cylinders and in the 2D limit 
of $L_y\rightarrow\infty$\cite{Zauner15}. We have not yet obtained a direct measurement of spin gap 
with iDMRG. However we can obtain easily the complete spectrum of possible \emph{correlation lengths},
by diagonalizing the MPS transfer operator. We set $\Lambda_{m,q}$ to be the largest eigenvalue of the transfer 
matrix after the identity for a wavefunction with number of states of $m$ and global spin of $q$. 
The correlation length, namely $\eta_{m,q}$, can be derived using $|\Lambda_{m,q}| = e^{u_0 / \eta_{m,q}}$, 
where $u_0$ is the size of the wavefunction unit cell in units of the lattice spacing. The 
principal correlation length is the \emph{largest} one of all $q$-sectors.

Hastings' 2004 theorem\cite{Hastings04} connect the size of the gap to an upper boundary 
for the correlation lengths of static correlation functions for a local, transitionally-invariant Hamiltonian 
(thus the size of the gap relates to the range of the correlations). For wavefunctions dominated by 
one-dimensional physics (built using the MPS ansatz), scaling behaviors are well known\cite{EntanglementScaling}. 
For critical, gapless states, the correlation length should diverge with increasing the number 
of states. These states have quasi-long-range (power-law) spin correlations, and diverging entanglement entropy.
In contrast, for gapped states, the correlation length should saturate toward 
a finite value with increasing $m$. For a 2D cylinder, the correlation length might diverge
with the cylinder circumference.

In \fref{fig:CorrLength}, we present our correlation length results for two quantum number sectors $q=0,1$. 
We note that for even-boundary topological sectors 
the correlation lengths for $q=0$ and $q=1$ cross at $m \approx m_{kink}$ (not shown 
in the figure). For fixed-width cylinders, looking to either individual values of $\eta_{m,0}$ 
or extrapolated value of the $q=0$ series at $m\rightarrow\infty$, namely $\eta_{\infty,0}$, one can 
observe that all topological sectors and system sizes have relatively small correlation length as 
$\eta \sim 1-10$ in units of the lattice spacing. This suggests that the correlation length remains 
finite in the 2D limit, however these extrapolations are very sensitive and although
the correlation lengths are rather small, for the larger system sizes the correlation length is
increasing with basis size in a way that is essentially indistinguishable from power-law (which is what
is expected for a \emph{gapless} system).
The existence of a gapped SL phase for the model is supported by previous 
DMRG studies\cite{Zhu-White_DMRG,Hu-Sheng_DMRG}, 
although this was later questioned by other numerical studies\cite{Iqbal16,Bishop15}.

\section{Scalar Chiral Order Parameter}
\label{sec:ChiralOP}

%%%%%%%%%%%%%%%%%%%%%%%%%%%%%%%%%%%%%%%%%%%%%%%%%%%%%%%%%%%%%%%%%%%%%%%%%%%
\begin{figure}
  \begin{center}
    \includegraphics[width=0.64\linewidth]{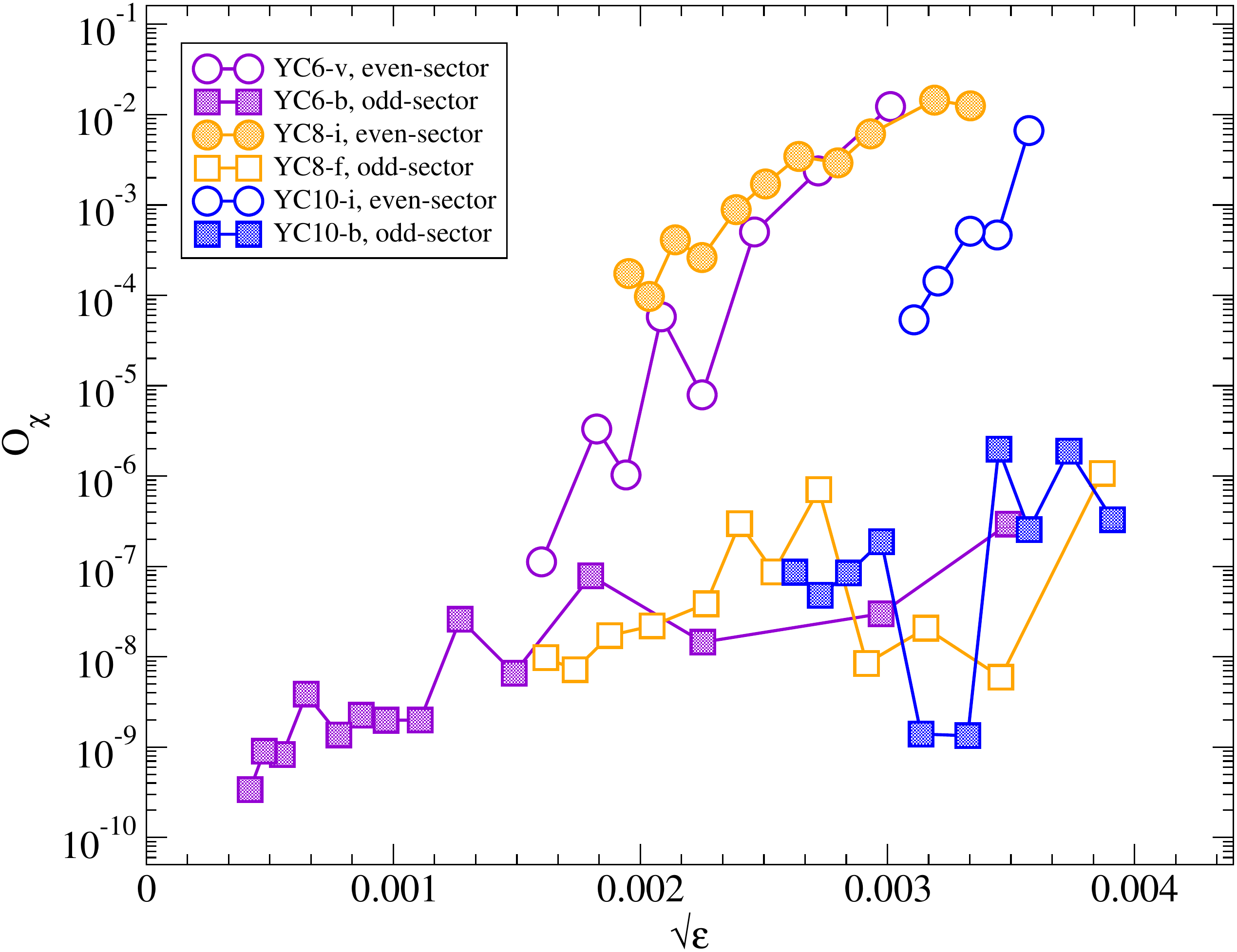}
    \caption{(Color online) SU(2)-symmetric iDMRG results for the scalar chiral 
    order parameter, $O_\chi$, against square root of truncation errors, 
    $\sqrt{\varepsilon}$ for different anyonic sectors and system sizes of the 
    THM at $J_2=0.125$. Results are presented in a semi-logarithmic
    scale to compensate for the rapid reduction of the order parameter in Y-axis.
    \label{fig:ChiralOP}}
  \end{center}
\end{figure}
%%%%%%%%%%%%%%%%%%%%%%%%%%%%%%%%%%%%%%%%%%%%%%%%%%%%%%%%%%%%%%%%%%%%%%%%%%%

In this section, we study the chirality of the spin-liquid phase. For spin-$\frac{1}{2}$ systems, the chirality 
also determines whether the state is planar or not\cite{Saadatmand15}. A commonly used observable to
measure the chirality is \emph{scalar chiral order parameter}, 
\begin{equation}
O_\chi = \frac{1}{u_0} \sum_{\la i,j,k \ra} (\vektor{S}_{i} \times \vektor{S}_{j}) . \vektor{S}_{k},
\end{equation}
Where $\la i,j,k \ra$ represent a nearest-neighbor triangular plaquette and the result is averaged
over the wavefunction unit-cell. A recent DMRG study\cite{Hu-Sheng_DMRG} found
long-range chiral ordering in the even-sector for even-width cylinders. 
We already note that there is \emph{no} sign of breaking of the time-reversal symmetry for either even- 
or odd-sectors (cf. main Letter). However to furthermore clarify the stability of chiral ground-states 
in the model, we measured $O_\chi$ for all of our obtained wavefunctions.

\fref{fig:ChiralOP} shows the detailed behavior of $O_\chi$ versus truncation ($\sqrt{\varepsilon}$).
The values of $O_\chi$ are very small, decrease rapidly with the truncation error, and extrapolate to
a value numerically indistinguishable from zero for $m \rightarrow \infty$. Thus in each case we find that
the state is \emph{non-chiral} and \emph{co-planar}. We note that for all 
available sizes of the even-sector, the wavefunctions have a small non-zero chirality for when keeping a
relatively small number of states, but this vanishes rather quickly with increasing basis size $m$.
In contrast, adding explicitly a chiral-symmetry breaking term ($J_\chi$) to the Hamiltonian readily breaks
the chiral symmetry\cite{Hu16,Wietek16} (while preserving $SU(2)$), supporting the notion that there is 
no spontaneous breaking of chiral symmetry in the ground-state, but a relatively small $J_\chi$ term 
is sufficient to induce a chiral spin liquid\cite{Wietek16}.

%\bibliography{mybib}

\cleardoublepage

\end{document}